# Homomorphic Encryption in Healthcare Industry Applications for Protecting Data Privacy

**J. S. Rauthan**

*Department of Computer Science and Engineering*

*G. B. Pant Institute of Engineering and Technology*

*Pauri Garhwal, 246194, India*

*jsrauthan1@gmail.com*

**Abstract.** Focussing on two different use cases—Quality Control (QC) methods in industrial contexts and Neural Network (NN) algorithms for healthcare diagnostics—this research investi-gates the inclusion of Fully Homomorphic Encryption (FHE) into real-world applications in the healthcare sector. We evaluate the performance, resource requirements, and viability of deploying FHE in these settings through extensive testing and analysis, highlighting the progress made in FHE tooling and the obstacles still facing addressing the gap between conceptual research and practical applications.

In order to show why FHE is the best technology for maintaining privacy in our applications, we start our research by describing the specific case study and trust model we're working with. We then concentrate on solving the bidirectional privacy difficulties and simplifying the way security is currently handled.

Choosing the two FHE frameworks most appropriate for industry development, we assess the resources and performance requirements for implementing each of the two FHE frameworks in the first scenario, Quality Control algorithms. Our understanding of the effects of optimizations is enhanced by this comparison, which also enables us to provide useful recommendations for the best FHE system and encryption method based on individual computing requirements. We further discuss the benefits of directly incorporating fundamental statistical functions into FHE libraries and the possibility of accelerating these procedures with dedicated hardware.

Regarding the second scenario, which involves NN models, we differentiate between two devel-opment approaches: converting pre-existing models into versions that are compatible with FHE

Address for correspondence: Department of Computer Science and Engineering, G. B. Pant Institute of Engineering and Technology, Pauri Garhwal, 246194, India

and creating models natively within an environment that is native to FHE. Although the former permits the application of tried-and-true NN architectures, it encounters difficulties and inefficien-cies in the translation procedure. On the other hand, the latter requires more sophisticated FHE tooling but offers streamlined development efforts and maybe improved accuracy. We illustrate the real difficulties in reaching high inference accuracy and reasonable runtime performance, par-ticularly for complicated models in healthcare diagnostics, through the prism of three NN models of different levels of complexity.

In conclusion, our findings demonstrate the effectiveness and resource consumption of the two use cases—complex NN models and simple QC algorithms—when implemented in an FHE setting. Although NN models, particularly those meant for production, suffer from significant runtime delay and development complexities in FHE scenarios, QC methods show great potential for inclusion into non-latency-critical FHE systems. We talk about how these difficulties affect FHE analysis and point out the main directions for further development in FHE methods, hardware approaches, and algorithm optimizations.

# 1. Introduction

In the current digital world, privacy-preserving solutions are crucial, particularly in the healthcare sector. Patient information must be protected by stringent data protection methods due to escalat-ing legislative obligations, such as the General Data Protection Regulation (GDPR) of the European Union. These rules mandate that healthcare organizations put robust safeguards in place to guarantee the privacy, confidentiality, and integrity of data. However, because to the requirement for scalable storage, processing power, and collaboration platforms, cloud computing is unavoidably incorporated into healthcare infrastructure, posing inherent privacy and security problems. Modern privacy pro-tection methods are necessary because of the interplay between regulatory compliance demands and the inherent potential risks associated with cloud computing. Thankfully, developments in the subject are moving from abstract ideas to functional, real-world approaches [1, 2]. The healthcare business needs these tools to traverse the complicated environment of digital privacy and security while utiliz-ing cloud computing, therefore advancements in the practicality and scalability of privacy solutions are crucial.

## 1.1. Problem Definition and Motivation

In order to meet the security and privacy standards that are fundamental to healthcare business op-erations, the article attempts to address several significant challenges in this area. Identifying and deploying efficient privacy-preserving solutions to address these requirements is the main objective of the study, which focuses on examining and clarifying the issue area.

The existing strategy taken by the healthcare industry to maintaining data confidentiality is the core of the issue. Current procedures place heavy operational costs even though they are meant to safeguard confidential data. A standard method that adds an extra degree of protection but may make data tracing and analysis more difficult is hashing IDs. Heavy logical and computational loads are also imposed by the protocols that are used to encrypt and re-encrypt data, along with intricate key distribution structures. These systems can be laborious and time-consuming, and they also involve user management and data access control. The complexity of procedures and resources is increased when external specialists are consulted to confirm the low danger of data disclosure. Overly intricate security protocols have the potential to introduce additional security flaws. Overall, despite their good intentions, these actions impair functioning, lower availability, and raise operational expenses.

Moreover, bidirectional privacy issues are a fundamental problem in the context of healthcare organizations' cooperation with outside medical laboratories. In particular, the interaction between medical laboratories that supply confidential patient data and cloud-based servers in the healthcare industry—which offer computing services—is where the issues lie. In this case, a privacy model is needed to safeguard patient data in the medical laboratory from access by the healthcare organizations, as well as the private algorithms and computational frameworks of the healthcare industries. With the classified nature of medical data and the highly competitive pharmaceutical research and development industry, achieving this two-way privacy is essential to protecting the confidentiality of both parties'operations. The intricate privacy requirements of the healthcare sectors' Collaborative Operations Model cannot be sufficiently addressed by traditional privacy solutions since they are frequently made for one-way privacy challenges.

Consequently, this paper's primary goal is to find and choose a suitable privacy solution that fits the particular situation and objectives mentioned above. Once a method has been selected, an extensive feasibility analysis will be carried out to assess how well it will work to meet the difficulties that have been identified. The research is concentrated on finding solutions that guarantee bidirectional privacy, streamline and improve the security procedures in place, and work in harmony with the computing infrastructure present in the healthcare sector. After that, the study will concentrate on putting into practice methods that give priority to raising performance and cutting down on resource consumption. As well as an estimate of the accompanying costs in terms of overhead, throughput, and other computer resources, this will give a primitive knowledge of how feasible current privacy-preserving solutions are in real-world commercial scenarios.

## 1.2. Existing Techniques Which Preserve Privacy

Following an explanation of the main issues and objectives, we now look into modern privacy-preserving technology. We will examine various existing privacy methods, show how they satisfy the many demands of current privacy issues, and evaluate whether or not they can help with our particular problems in the parts that follow.

### 1.2.1. Differential Privacy (DP)

Users within a dataset are intended to have high privacy protections thanks to Differential Privacy (DP) [3]. It makes it difficult to deduce any individual's details from the aggregated data by ensuring that the addition or deletion of a single individual's dataset does not materially alter the results of any analysis. [8] Without affecting the privacy of specific patients, a health organization might use DP to publish aggregate information concerning patient outcomes.

### 1.2.2. Secure Multi-Party Computation (SMPC)

Using a cryptographic technique called Secure Multi-Party Computation (SMPC), several people can work together to calculate their private data without disclosing it to outside parties or to one another [4]. It works well in situations when several people must collaborate to analyze data. [5] In collabo-rative studies, like in the medical industry, a common use case is when various hospitals want to work together to analyze personal data to enhance treatment procedures, but they don't want to share private patient data.

### 1.2.3. Zero-Knowledge Proofs (ZKPs)

Through the use of zero-knowledge proofs, one party (the prover) can demonstrate to another (the verifier) that a specific assertion is true without disclosing any details other than the statement's inher-ent validity. This makes it possible for the prover to remain anonymous while still giving the verifier confidence. ZKPs can be utilized for validating payments without revealing the payment's specifics to the entire network, which is a typical use scenario in blockchain and cryptocurrency transactions [6].

### 1.2.4. Trusted Execution Environment (TEE)

A computer's central processor unit (CPU) has a safeguarded section called a Trusted Execution Envi-ronment (TEE) that is intended to keep confidential data and algorithms safe from access or alteration by unauthorized activities. Critical operations can take place in a "safe space" created by a TEE, which isolates data and programs inside a private enclave within the main CPU. In cloud computing, TEEs can be utilized to offer secure operating environments, protecting sensitive activities from users and various cloud services [7].

### 1.2.5. Homomorphic Encryption (HE)

One technique of encryption known as homomorphic encryption (HE) [8] that permits calculations to be done on encrypted data without the need to first decipher it. Only the private key holder is able to decipher the computations' outcomes, which are still encrypted. Individuals can encrypt and upload confidential information to the cloud via cloud computing, which is a common application for HE. After then, the encrypted data can be used by the cloud service to do calculations (like statistical analysis) without ever exposing the original data. For sensitive information such as medical records, this is especially crucial [9, 10].

These privacy-preserving techniques have inherent limits despite their many benefits. It is difficult to strike a balance between data utility and privacy when using DP, for example, since the noise introduced to protect privacy can reduce the usefulness of the data. Because SMPC involves a lot of processing and a lot of communication overhead between participants, it might not be appropriate for real-time or large-scale applications. ZKPs are effective for protecting privacy, but they are difficult to create and frequently need a lot of processing power to produce and validate, which limits their scalability and usage. TEEs offer a safe environment for data processing, but their reliability mostly depends on the computer's hardware. Finally, while HE enables reliable computation on encrypted information, its current limitations in terms of computational overhead and poorer performance when compared to plain computations may limit its use for applications that require a lot of resources or are time-sensitive. While many of these technologies are still developing, these constraints frequently limit how effective they may be. It necessitates serious thought and weighing trade-offs in real-world situations as a result.

## 1.3. Why Use Homomorphic Encryption?

It has been determined that HE is the best suitable privacy-preserving solution when considering the unique demands and obstacles faced by the healthcare industry. HE is unique in that it enables com-putations on encrypted data without disclosing the actual data to any of the parties involved [11, 12]. In order to protect patient confidentiality, the healthcare industry can now handle and investigate en-crypted patient data that it receives from laboratories without having access to the original data. Addi-tionally, because they only communicate with and transfer data in an encrypted manner, laboratories operating in a server-client situation lack access to the secret algorithms and models used by the health-care industry. Since this satisfies the requirements for both directions' privacy, HE is the best option in this case.

Furthermore, HE can streamline the scenarios involving the confidentiality management systems now in place in the healthcare industry. Traditionally, preserving data confidentiality has required a number of intricate procedures, including data anonymity, encryption/decryption phases, stringent access controls, and thorough audit trails—especially in highly restricted situations. Despite being essential for security, these procedures can entail high operating complexities and requirements for resources. By performing calculations on encrypted data, HE shifts the paradigm by eliminating the requirement for repeated encryption and decryption phases. HE eliminates complexities and vulner-ability caused by various public key infrastructures and repetitive re-encryption by keeping data en-crypted throughout the process of computation. This capacity greatly lowers the amount of resources needed and the complexity of operations.

Although alternative privacy solutions have been explored, none of them fully addresses the spe-cific requirements of the research [8]. In statistical data sets, for instance, DP can efficiently guarantee anonymity, but it is unable to satisfy the needs of two-way privacy. When numerous parties and their corresponding information sets are involved in collaboration and calculations, SMPC is a more appropriate method. Similarly, ZKP offers robust privacy guarantees, however it works better for ver-ification procedures than for analyzing data. While TEE offers a secure processing environment that can help with the bidirectional privacy barrier, it is mostly dependent on hardware integrity and is unable to help lower the total complexity of the security handling process in the healthcare industry.

As a result, HE provides a complete solution that satisfies the need for two-way privacy and can assist in simplifying the way that privacy is now handled.

## 2. Homomorphic Encryption and its Associated Tools

Homomorphic Encryption (HE) can be subsequently classified as Partial Homomorphic Encryption (PHE), Some-what Homomorphic Encryption (SHE), and Fully Homomorphic Encryption (FHE) based on the range and complexity of operations permitted on encrypted data. Among these, SHE enables a certain amount of both addition and multiplication operations, but not sufficient to do arbi-trary operations, whereas PHE only permits one kind of operation (either addition or multiplication). On the other hand, FHE is the most advanced and useful subcategory since it permits an infinite de-gree of both operations, allowing for extensive and complex computation without affecting privacy [13, 14].

SHE's capability is congruent with certain industrial application scenarios where a small number of multiplications may be necessary. [15] Still, a lot of sectors opt to adopt FHE in certain situations. FHE's scalability and flexibility are major contributors to this conclusion. Business demands and data processing capabilities may change and surpass SHE's constrained computational scope, even though SHE might be adequate for existing applications. More operations can be implemented using FHE, making it a more versatile solution that doesn't need to be reassessed as requirements change. Further, by using FHE from the beginning, system architectural complexity can be decreased by standardizing encryption procedures across various applications within an enterprise and streamlining security mea-sures. This makes maintenance and updates simpler, speeds up the development process, and lowers the chance of security flaws resulting from inconsistent systems.

Nevertheless, decreased productivity and more computational overhead are trade-offs associated with FHE. Costs would rise as a consequence, and more powerful computer resources would be needed. However, because of its adaptability and the long-term advantages of having a scalable en-cryption scheme that doesn't require continual adaptations to meet new or developing demands, the investment in FHE is justified.

## 2.1. Fully Homomorphic Encryption (FHE)

Since its conception, the idea behind FHE has undergone substantial development. Craig Gentry took the initiative in creating the first revolutionary FHE solution in 2009, which set the stage for fur-ther advancements in the industry. This groundbreaking work—often referred to as first-generation FHE—introduces the idea of bootstrapping to refresh ciphertext and takes advantage of perfect lattices, allowing for limitless computation on encrypted data. Second generation FHE methods followed, which were more efficient and readily available. These include the Cheon-Kim-Kim-Song (CKKS), Brakerski-Fan-Vercauteren (BFV), and Brakerski-Gentry-Vaikuntanathan (BGV) schemes, which im-prove the usefulness of FHE and solve some of the performance problems with Gentry's original structure. Research is currently moving toward the third generation, which is based on the Gentry-Sahai-Waters (GSW) method and consists mostly of the CGGI (Chillotti-Gama- Georgieva-Izabache, or TFHE) scheme. BFV and BGV, two second-generation systems, require minutes to complete, but this generation, which focuses on quick bootstrapping, achieves performance in less than 0.1 seconds. [16] Along with this, current research aims to improve security settings, decrease computing cost, and create more intuitive implementation resources and libraries. Growing practical applicability of FHE is shown in its progression from an abstract framework to an actual tool.

Every contemporary FHE technique is created to satisfy various requirements for computing and data processing. The foundation of the BGV and BFV methods is lattice-based cryptography. They allow a certain number of arithmetic operations to be performed on encrypted data, with a concen-tration on integer arithmetic. However, the CKKS technique is appropriate for machine learning and data analysis since it permits computations on encrypted floating point integers. The CKKS method, on the other hand, is approximate by nature, giving up some precision in exchange for the capacity to handle complicated computations. In contrast, the BGV and BFV schemes are designed for accurate arithmetic on encrypted integers, yielding precise outputs. Since the CGGI technique was designed mainly with binary operations in mind, it is particularly well-suited for logic-based calculations and binary arithmetic. Significantly, further study has increased its capacity to include arithmetic circuits, increasing its applicability to a larger variety of calculations.

## 2.2. Fully Homomorphic Encryption Tools for Industry Development

Technological advancement greatly depends on organized libraries and tools, particularly for industrial productivity. Although there are occasionally subtle differences between them, these tools are often classified as high-level compilers and low-level libraries in the context of FHE [16]. In this study, we will consistently refer to both kinds of resources as ”tools” for the sake of clarity. These technologies are intended to make the process of developing apps simpler by handling cryptographic primitives internally and offering data analysts, programmers, and security professionals intuitive interfaces. Following an extensive examination of scholarly works and internet sources, we discovered a wide range of FHE resources. Within the context of this study, a couple of these jump out as being especially intriguing for industrial advancement:

### 2.2.1. OpenFHE

OpenFHE, an open-source library created by a consortium of research organizations and industry pro-fessionals, is the replacement for PALISADE. Its goal is to provide an all-inclusive, flexible, and intu-itive FHE toolbox. The Rust API is continuously being developed, however the C++ and Python APIs are already accessible. All popular FHE methods are supported by OpenFHE in an efficient manner, including CKKS for real number arithmetic, FHEW and CGGI/TFHE for boolean circuits and arbi-trary function assessment, and BFV and BGV for integer arithmetic. Particularly, multiparty variations and proxy re-encryption for particular techniques are supported by OpenFHE, along with the ability to switch between CKKS and FHEW/TFHE to assess non-smooth functions like comparison. In [17] To help users with advancement, the library offers thorough documentation and demonstrations.

### 2.2.2. TFHE-rs

ZAMA created TFHE-rs, a Rust implementation of the TFHE/CGGI protocol enabling safe integer and Boolean arithmetic operations on encrypted information. To accommodate various development demands, it provides a client-side WebAssembly (WASM) API in addition to Rust and C APIs. By removing the complexity of low-level cryptography processes, TFHE-rs aims to offer a reliable, simple, high-performance approach that is prepared for industrial use. [18] There is published demonstrations and manuals.

### 2.2.3. Concrete-ML

Concrete ML, another open-source platform created by ZAMA, uses FHE to guarantee machine learn-ing that protects privacy. [19] It is intended to allow data scientists to experiment with FHE interop-erable machine learning algorithms regardless of their experience with cryptography, thanks to the availability of Python API. This may interconnect with several model variations, for example linear models, tree-based models, and neural networks, and offers reliable analysis on encrypted data using well-known programming techniques from packages such as scikit-learn and PyTorch. There is also provided demonstrations and manuals.

Numerous other FHE methods have been taken into consideration, including SEAL, HElib, EVA, nGraph-HE, SEALion, and many more. In [16] For varying cryptographic requirements, each offers special features and potent powers. But there are drawbacks when you match the requirements of our research with these technologies. For instance, using tools like SEAL may need a thorough grasp of cryptography, which could extend the time and complexity needed for industrial development. Certain computations that are crucial to our analysis are not supported by some tools, like EVA. Compatibility with the organization's standards may be difficult with certain tools, such as SEALion, because they are not open-source. Concerns over long-term assistance are further raised by the fact that some tools, like nGraph-HE, are no longer supported or have little public support.

# 3. Healthcare Use Cases and Examination

We now turn our attention to the practical use of FHE in this paper's efforts to solve privacy concerns related to health information. The reasons for choosing particular FHE tools for particular use cases will be discussed in this section, along with the experimental setup that was utilized to incorporate FHE into the healthcare apps. In addition, specific instances where FHE is applicable and crucial will be explored. The section that follows will provide the analysis and findings of the experiment, analyzed the resource demands and performance trade-offs, and shed light on whether FHE approaches are feasible in real-world healthcare scenarios.

## 3.1. Scenario and Trust Model

As previously stated, FHE results in considerable resource requirements and computational overhead. Therefore, before deciding whether FHE is practicable for a certain application, it is crucial to carefully consider various circumstances as other, less sophisticated and less expensive methods can be adequate to meet privacy concerns.

As figure 1 illustrates, we provide a client-server bidirectional privacy case in our environment. The customer is on behalf of medical data providers, including lab and research facility equipment that contains private patient data that must be strictly protected. In contrast, a server running on an edge or cloud computing architecture serves as a data processor. It analyzes patient data using exclusive algorithms or models from the healthcare sector and then sends back the analyzed outcomes.

In the present scenario, the client is responsible for protecting the privacy and confidentiality of the medical data from any potential security lapses or illegal access within the server ecosystem. In the meantime, the server under the authority of the healthcare sector needs to shield the client from seeing its exclusive models and algorithms. Additionally, it's important to note that the server is always set up in trusted zones for the healthcare industry, thus extra security against hardware or side-channel attacks doesn't need to be taken into account. The requirements for two-way privacy are evident in this case scenario: not only must the server's calculation be hidden from the client, but the client's data must remain private during processing. Due to FHE's capacity to process encrypted data and preserve both parties' security and privacy demands without sacrificing the data's usefulness, using it for privacy protection is a case study for privacy management.

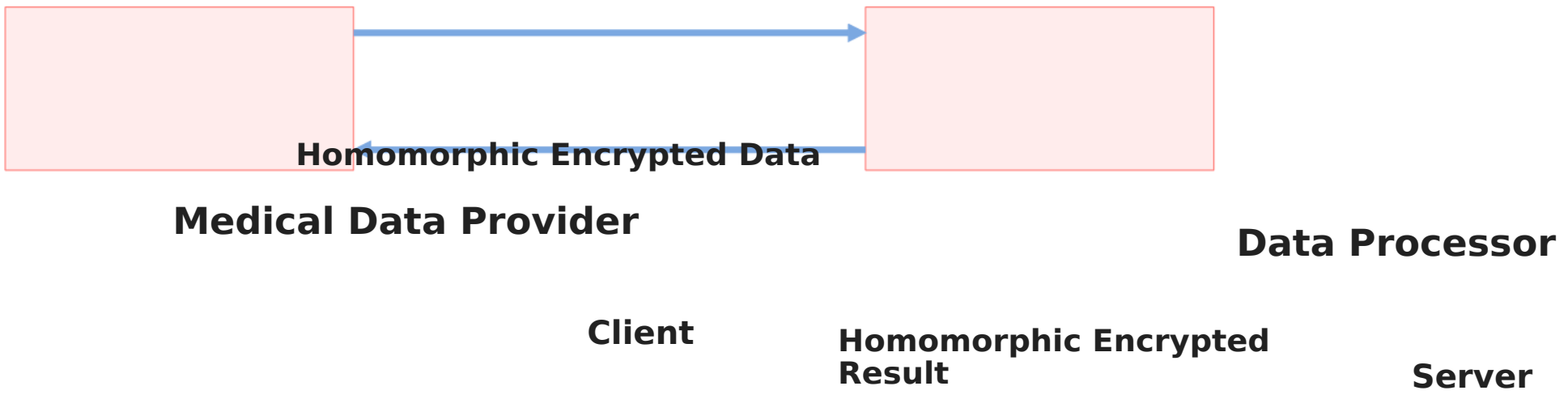

Figure 1. Client-Server Bidirectional Privacy Case .

Numerous applications and calculations can be carried out by a client and server in the context of the scenario and trust model that have been presented. The goal of this paper was to evaluate the viability of implementing FHE in two specific real-world applications: rule-based quality control and neural network models, which are significant healthcare use cases. Because of their various computing requirements and significance to the medical field, these instances were chosen.

## 3.2. Use Case 1: Quality Control Based on Rules

Within the healthcare sector, Quality Control (QC) is a procedure used to guarantee the accuracy and dependability of laboratory results, as well as the proper observance of regulations by medical practices and equipment functionality. It aids in preventing therapeutic errors, ensuring that patients receive excellent care, and keeping patients secure.

In this study, we focus on rule-based quality control (QC), which acts as an investigator over the integrity and validity of data. Westgard Rules are applied in rule-based quality control [20]. The foun-dation of these guidelines is statistical evaluation, which is essential for assessing and guaranteeing the reliability of laboratory outcomes. Every rule is the result of several computations put together.

James Westgard, the creator of Westgard Rules, explains how multi-rule QC, or Westgard Rules, can assist individuals detect anomalies in a sequence of data in this straightforward way: ”My daughter Kristin liked to party when she was younger and living at home. When she told me she was going to be out late again one day, I felt compelled to impose some parental control over her schedule. I therefore warned her that she was going to get into serious trouble if she left after three, twice after two, or four times after one. We call that multirule control [20].

Comparably, in this use instance, a set of rules including fundamental statistical computations like mean, variance, and other pertinent metrics will likewise be applied to the patient data.

| Rule | Input | Outp |
|---|---|---|
| First Rule | Decimal | One |
| Second Rule | Decimal | One |
| Third Rule | Decimal | One De |

Table 1. The QC Rules' Data Types.

### 3.2.1. Selection of Rules in Quality Control

The analysis selected three of the most typical QC rules for a feasibility assessment out of the more than twenty that are already accessible in production applications. While allowing for the potential for in-depth research, this selection aims to offer a thorough overview.

### 3.2.2. Program Evaluation

It is essential to carry out a program analysis in order to comprehend the precise requirements prior to choosing an FHE development tool. This study aids in identifying the application's needs for output accuracy, data type, and computations. All of these elements work together to influence the choice of appropriate FHE systems.

- **Data Type:** The data type of the chosen QC rules' input and output is displayed in Table 1 below. Observe that every decimal represented in this case is in the format XX.XX, meaning that it has two decimal places of precision and ranges from 0 to 100.

- **The Required Computations:** The calculations used in the chosen QC guidelines are displayed in Table 2 below. The computations in this use case are notable for their heavy reliance on non-smooth/binary operations like comparisons and logic gates. Additionally, two of the rules call for arithmetic computations. This is in stark contrast to a lot of recent HE research, particularly that which has an industrial background, where the capabilities of certain HE schemes or tools are typically taken into consideration when selecting a test algorithm. In such an academic setting, the experiment tends to select algorithms that exclusively contain arithmetic operations, and vice versa for binary operations, if the target HE systems or tools mostly enable arithmetic computations.

  On the other hand, because this work is grounded in a practical application inside an actual context, it is necessary to use tools that are capable of effectively handling binary and arithmetic computations as needed. Rather than being constrained by the theoretical breadth of current re-search and isolated academic frameworks, this method makes sure that the solutions we develop are effectively applicable to and advantageous for real-world healthcare situations. Therefore, selecting and developing tools presents special hurdles because of this computing requirement.

- **Standards for Output Accuracy :** The outcomes of QC regulations are critical in the health-care industry since they determine the accuracy and dependability of the goal data and the ex-perimental procedure. Consequently, we anticipate that these rule outputs will have the highest level of accuracy feasible. Although a slight error rate could be acceptable and can be mini-mized by performing the calculations multiple times, the outcome should preferably be exact rather than approximative. Because of this high standard of accuracy, judgments on healthcare based on these results are guaranteed to be based on the most trustworthy data available.

| Rule | Arithmetic Calculations | Binary Calculations |
| --- | --- | --- |
| First Rule | None | Comparison, Logic Gates |
| Second Rule | Addition, Subtraction, | Comparison, Absolute |
| Third Rule | Addition, Subtraction, | Comparison, Logic Gates, |

Table 2. The Calculations Associated with the QC

### 3.3. Use Case 2: Models of Neural Networks

In contrast, neural network (NN) models are a more sophisticated computational model that can take advantage of FHE’s features. NN models are essential to improving therapeutic and diagnostic proce-dures in the healthcare sector. They specifically aid in the classification of diseases based on genetic information or symptoms in classification issues. When it comes to image processing, these models considerably improve the speed and accuracy of interpretation when compared to manual techniques. They do this by analyzing medical pictures, such as X-rays, MRIs, and CT scans, to find anomalies, tumors, fractures, or disorders. NN models can aid in the fast interpretation of complex genetic data and the very accurate detection of infections during PCR testing. Customized medicine and focused treatment approaches are made possible by NN models, which are also used in genetic testing. These models evaluate gene sequences to uncover markers linked to certain illnesses or disease predisposi-tions. In general, neural network models are transforming the healthcare sector by offering a wide range of applications that are quicker and more precise diagnostic and prognostic tools.

We are able to conduct advanced analysis and forecasts on encrypted medical data using NN models, all while maintaining patient confidentiality. Since deep learning and artificial intelligence (AI) are becoming increasingly important in medical research and examinations, this technique is very promising.

Three neural network models, varying in complexity and capability, are the subject of this research. The first is a simple classification model intended for use in medical diagnostics. The model is a typical use case in HE research, despite the fact that its simplicity has prevented it from being employed in a production setting. Beginning with this fundamental model, our goal is to assess how well the available tools can handle fundamental neural network models and assess resource consumption and performance in this kind of environment.

The second model is made for classification and image processing. With tens of thousands of pa-rameters and hundreds of layers, it is a fully connected neural network model. Although this particular model has not yet been put into use, models with a comparable structure and number of layers are be-ing actively utilized in real-world applications, especially for imaging-based tumor and other medical anomaly detection. We are able to investigate the usefulness of HE methods for image processing and classification with this model.

And finally, the third model, which is the most complex of the three, is made especially for PCR tests. Presently, this sophisticated model is employed in real-world settings for routine healthcare procedures. This model presents major problems for HE research with its complicated structure, tens of thousands of parameters, and more than 40 layers. The goal of the research is to push the limits of what is currently possible with HE in an industry perspective by including this model into our study, especially in the healthcare sector where speed and accuracy are critical.

# 4. Evaluation Based on Experiment

In this research, we undertake experimental assessments throughout specified QC algorithms and NN models to examine the applicability of specific FHE technologies in real-world healthcare environ-ments. We give particular attention to both performance during runtime and utilization of resources in the benchmark assessments for a number of important reasons:

- **Efficiency during program execution:** Because the daily activities of the healthcare industry consist of handling hundreds of thousands or even millions of samples, runtime efficiency plays a crucial role in processing extensive amounts of data quickly and guaranteeing timely delivery of findings.

- **Client-side utilization of resources:** Portable medical tools are frequently used in healthcare environments, and these devices might not have as much memory or processing capability as a typical laptop or server processors. In order to guarantee that the FHE tools we select for our studies can be used on these portable machines, it is crucial that they be resource-efficient.

- **Costs associated with transferring data:** The global nature of the healthcare industry means that network transmission costs can greatly impact latency due to the volume of incoming and outgoing data transfers. Hence, in our FHE applications, we investigate the size of inputs and outputs.

## 4.1. Selection of Tool for Full Homomorphic Encryption

1. **Necessary conditions and limitations** We give consideration to the constraints and require-ments arising from the technical implementation as well as the industry context while choosing FHE tools. These elements are crucial when taking into account industrial needs:

- **Compliance with security and privacy regulations:** Ensuring compliance with stringent security and privacy standards, such as upholding a minimum 128-bit security level, takes precedence.
- **Source code and reliability:** In general, industry prefers open-source technologies or those made by reputable outside organizations to guarantee dependability and transparency. Regular maintenance should also be done on code bases.

- **Ease of use:** For non-security specialists like software engineers and data scientists, the tool should be easy to use. This will reduce the learning curve and encourage wider indus-try utilization.
- **Functionality and interoperability:** Tools should ideally support programming languages that are currently widely used in the industry's development ecosystem and demonstrate efficient run-time and memory consumption.

Regarding technical advancement, we require the following in light of our examination of the QC algorithms and NN models involved:

- **Dealing with decimals:** It should be possible for the chosen FHE tools to handle decimal numbers.
- **Assistance in calculating QC algorithms:** The tools chosen for the
- **Assistance for neural network models:** The tools used must support all of the associated operators, including Sigmoid, Relu, and Flatten, for the NN model application scenario.

2. **Chosen Tools** We chose the following FHE instruments for our paper in accordance with the specified specifications and limitations:

- **In the scenario of quality control (QC):** We choose TFHE-rs and OpenFHE. These tools meet our security and privacy requirements, are open-source, and can perform the binary and arithmetic operations needed for this particular scenario. Because of third-generation research, the TFHE(CGGI) method, on which TFHE-rs is based, is able to perform both types of computations. Although CKKS and FHEW/TFHE provide binary operations and arithmetic operations, respectively, OpenFHE facilitates scheme switching between them. Their stated performance traits and compatibility with current development frameworks also played a role in the selection process.
- **In the scenario of utilizing the NN model:** Concrete-ML was chosen for the model scenario of the NN. While a lot of other tools can handle basic NN models and low-level machine learning (ML) activities, this tool has a unique feature that makes it stand out: it offers a high-level API that mimics the ML/NN native environment in a user-friendly manner suitable for data analysts and developers. We find that it is especially well-suited for handling intricate, high-volume NN models in our paper because of this support.

### 4.2. Setup for the experiment

- **Algorithms and Models Tested:** As previously mentioned, our experimental assessment en-compasses two primary use cases: NN models, which are intended for image classification and PCR tests, and rule-based QC algorithms, which utilize statistics to assess the precision and dependability of medical data.

- **Hardware Setup:** The following specifications applied to a standard hardware that was used for the research. CPU: Intel Core i7 processor, which has 16 logical threads and 8 cores; 32 GB of memory, Additional details: The CPU has a base clock of 3.8GHz and a maximum turbo frequency of 5.1GHz. It has a 16MB L3 cache installed.
- **Software Setup:** Every experiment was carried out utilizing the Ubuntu operating system in a Linux-based environment. Regarding programming languages, we created implementations in Python for Concrete-ML, Rust for TFHE-rs, and C++ for OpenFHE.

## 4.3. Benchmark and Assessment

The experimental findings regarding resource use and runtime performance are shown in this section. Additionally, using development process difficulties and benchmarks, it assesses the viability of FHE in the chosen healthcare scenarios.

1. **Use Case 1: Quality Control Based on Rules**

   Three distinct QC algorithms, or rules, each exhibiting varying degrees of effectiveness and complexity, are examined in this use case.

   - **Rule 1: Fundamental Decimal Binary Operations**
     Among the tested QC methods, this one is the most basic. It primarily carries out a se-quence of binary operations. An array of decimal integers is the rule's input, and its output is a single bit that can be represented as either 0 or 1, denoting the pass or fail status of the QC. This rule demonstrates a fundamental quality control procedure in which choices are made using straightforward yes/no standards.

     The ability to process several inputs simultaneously through batching improves runtime speed, which is a major benefit of employing OpenFHE. In particular, when analyzing the execution time with different batch sizes, this efficiency is clearly proved in our tests. Figure 2 shows that when batch sizes increase, there is a noticeable drop in the total run-time for an individual batch of executions. Each input's average runtime within a batch is displayed concurrently in figure 3. For encryption and decryption on the client side and server side calculation, batching is undoubtedly advantageous for efficiency.

     It is noteworthy to emphasize that the advantages of batching plateau after a given amount of time. According to our research, a medium batch size—64 or 128 in our example—is frequently adequate to drastically cut down on runtime. Beyond this range, increasing the batch size yields very slight gains in performance. This finding points

to an ideal range for batch sizes that strikes a compromise between efficiency increases and real-world factors like memory consumption and complicated data management. This knowledge is especially helpful for applications where resource usage and performance are important factors.

The effectiveness of TFHE-rs deployments in different setups and configurations is dis-played in Figure 4. Since TFHE-rs is based on the TFHE (CGGI) system, handling deci-mal input directly is not something it can do by default. Our method entails converting the

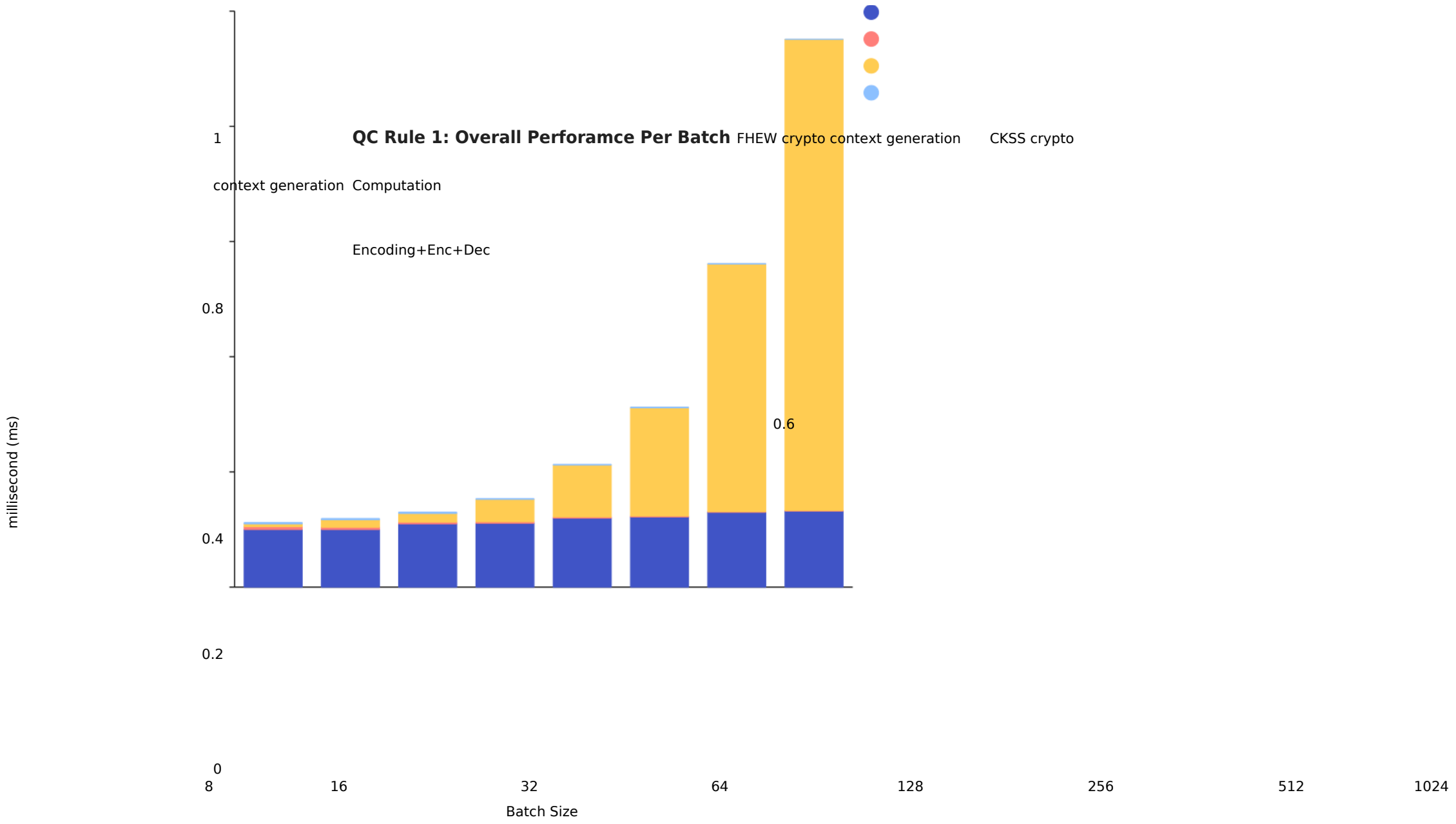

Figure 2. Implementations of OpenFHE for the First QC Rule, Efficiency for a Single Batch of Input.

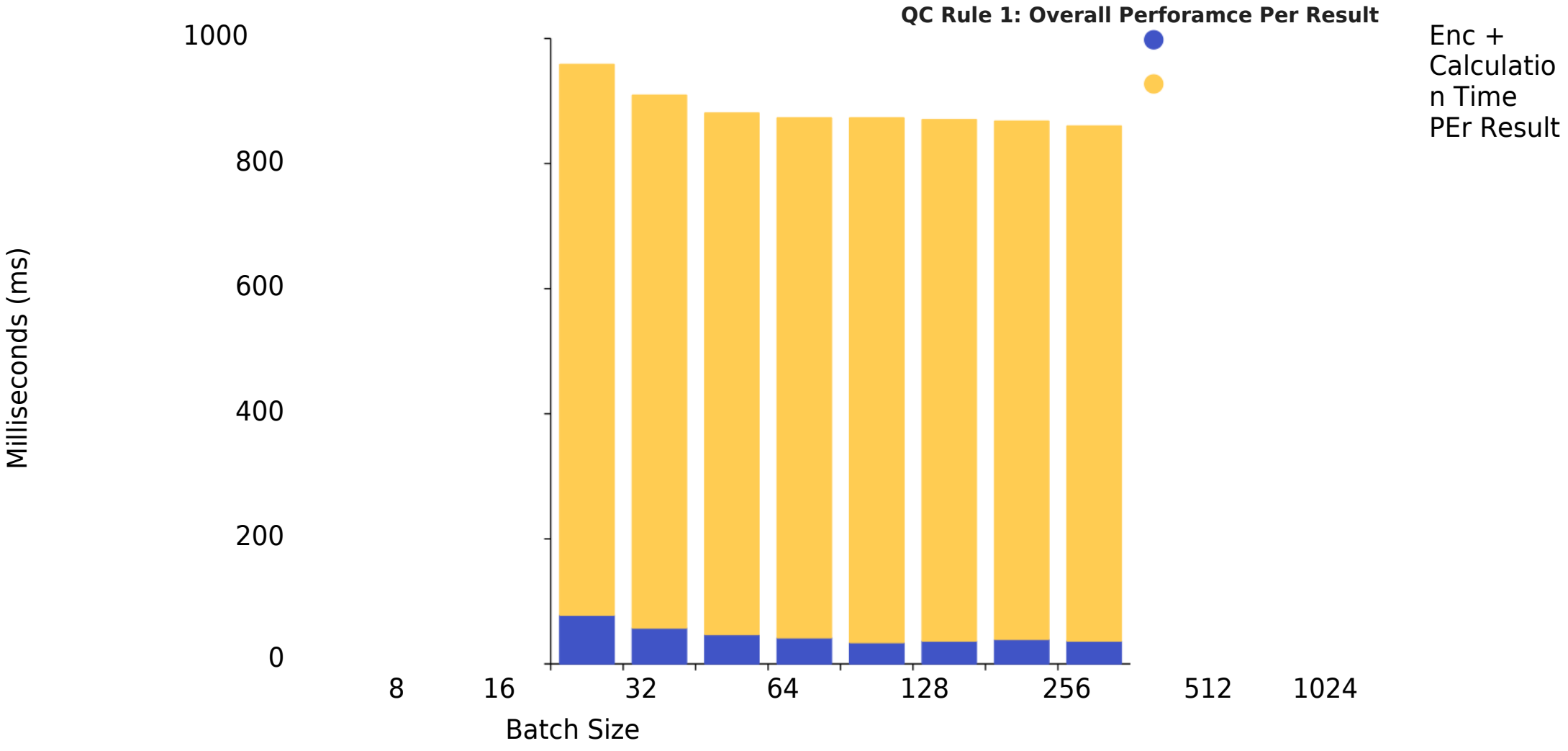

Figure 3. Batch Implementations of OpenFHE for the First QC Rule, Average Performance for Each

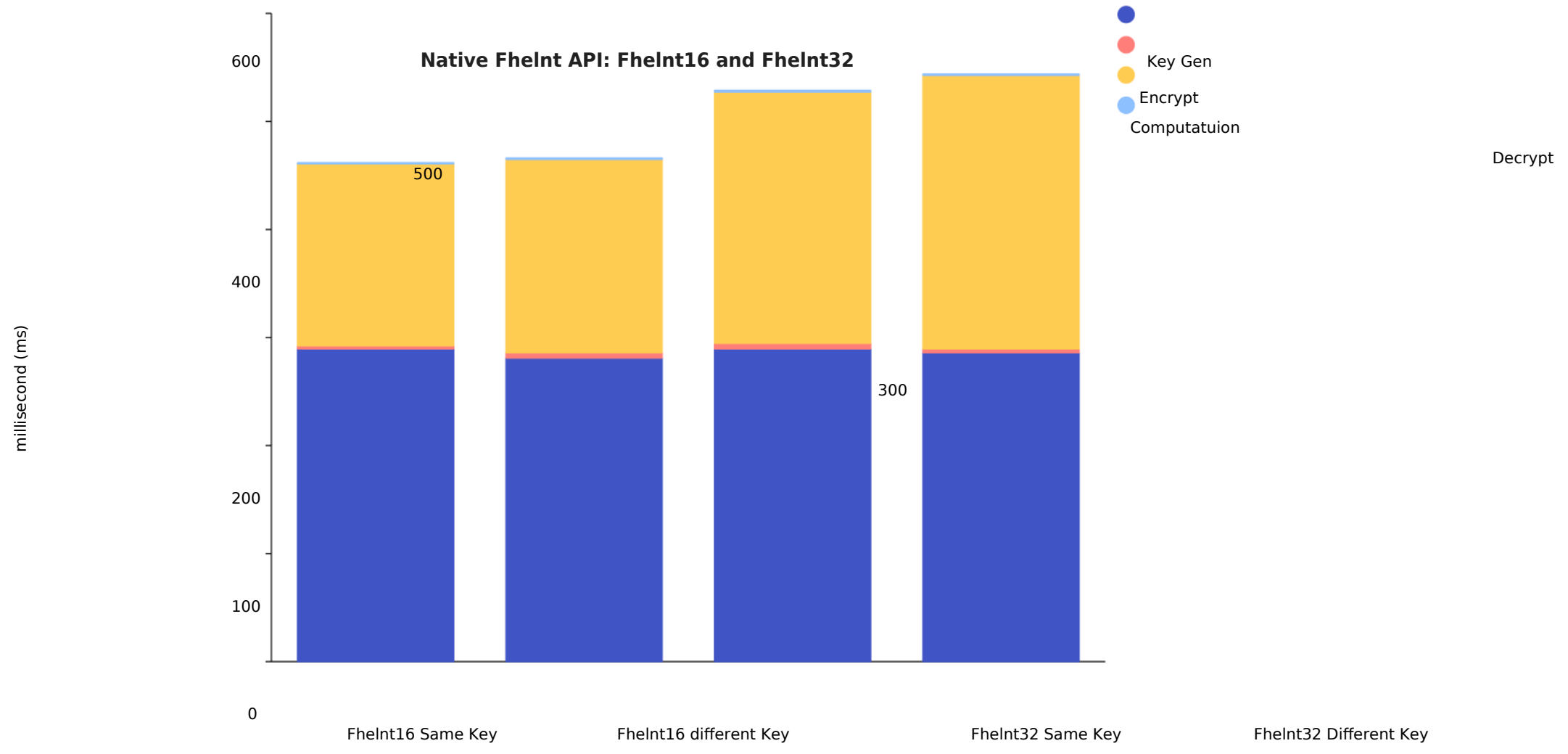

Figure 4. setups. Implementations of TFHE-rs for the First QC Rule, Performance Under Various configurations and raw decimal inputs into integers by multiplying them by a constant in order to get around this restriction. By doing this, we can take use of the scheme's capacity to handle integers up to 512 bits and represent decimals appropriately. We used integer sizes of 16 and 32 bits in our research. While 32 bit integers are selected to meet future requirements requir-ing larger numbers or greater accuracy decimal representations, 16 bit integers are enough for the current environment.

In addition, we provide two separate settings to represent the operational variations be-tween the client and the server: the server can deal with input sets from multiple clients, requiring frequent key changes, whereas the client may handle millions of input sets under a single stable key pair. The ”same key” and ”diff key” settings for these cases are marked in the figure 4. Significantly, the benchmark results we obtained indicated minimal dif-ferences in runtime between the two setups; hence, we disregarded this distinction in our further research.

Figure 6 shows that TFHE-rs performs better than OpenFHE on daily tasks including client side encryption and decryption and server side calculation when compared the most optimal solutions between OpenFHE and TFHE-rs. The difference becomes even more pronounced for infrequently conducted key generation and crypto-context setup, as figure 5 illustrates.

It is clear what is causing this discrepancy at its core: In order to facilitate binary op-erations, OpenFHE first accepts inputs using the CKKS method before switching to the FHEW/TFHE scheme for binary processing. This scheme change requires a lot of re-sources; more cryptographic context must be generated by the client, and more computa-tional processes must be carried out by the server.

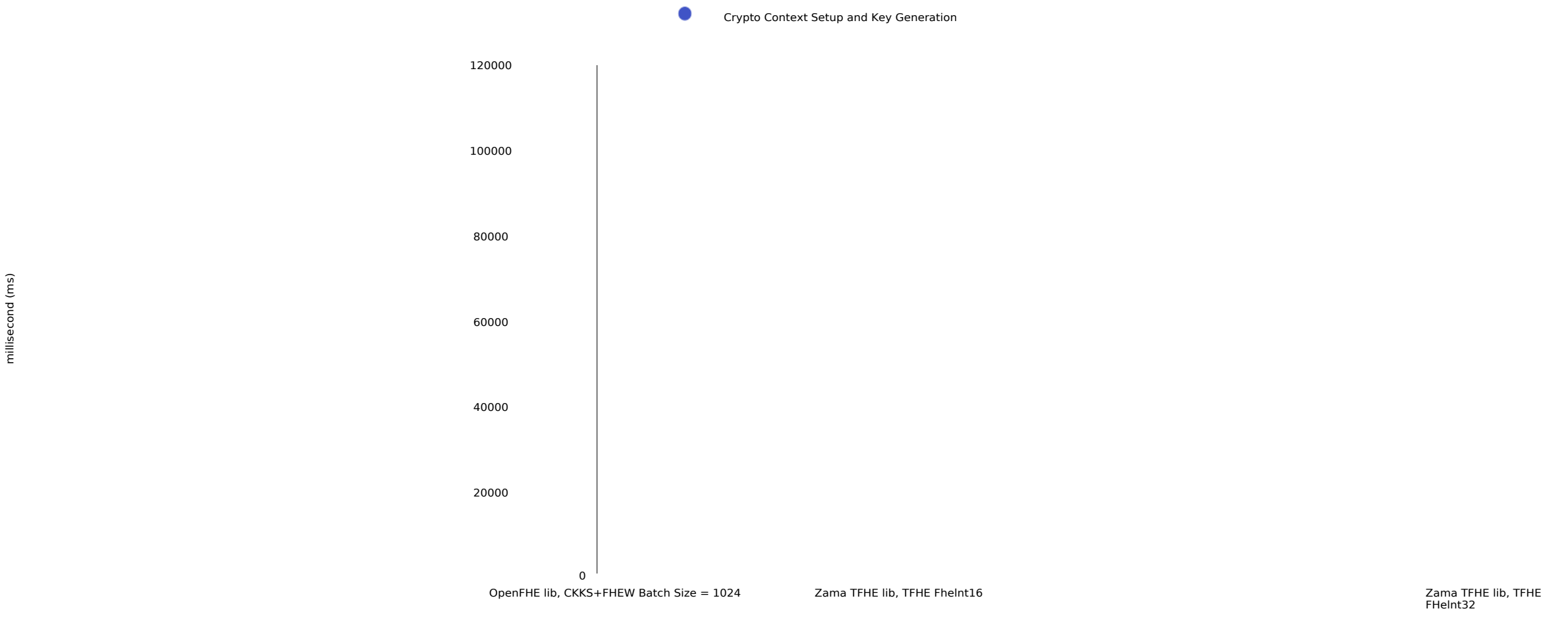

Figure 5. Comparing the top implementations of the initial QC rule in terms of key graphic context between OpenFHE and TFHE-rs.

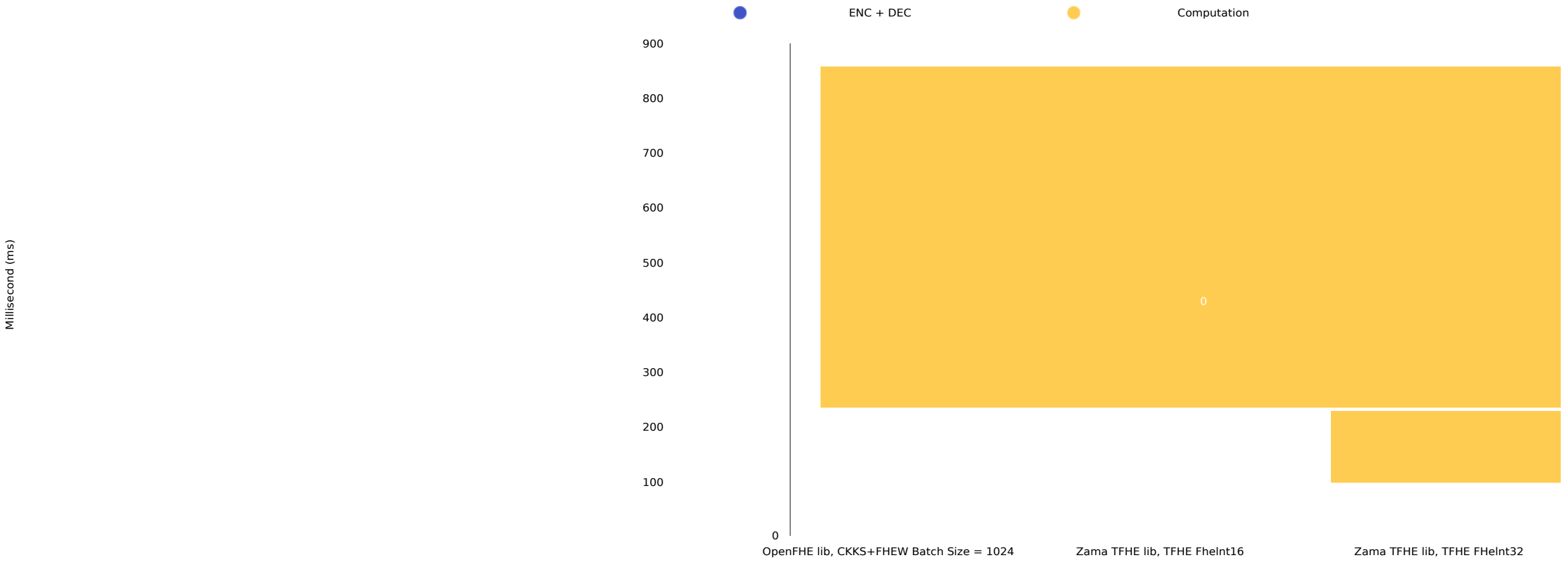

Figure 6. Comparison of top implementations of the initial QC rule in OpenFHE daily operational performance in encryption, decryption, and computation.

In contrast, TFHE-rs is designed for binary operations and is largely based on the TFHE (CGGI) technique. With the help of this specialization, TFHE-rs can perform binary op-erations more skillfully and organically, negating the need for scheme switching or other setup procedures. The direct support of binary operations by TFHE-rs results in significant performance gains, particularly in applications like this QC rule where binary computation is common.

The significance of choosing tools that closely match the unique computational require-ments of a paper is highlighted by this efficiency, underscoring the crucial role that tool selection plays in maximizing the performance of real-world applications.

- **Rule 2: Combination of arithmetic and binary operations**

  This rule takes an array of decimals as inputs, much like the last one. But it does more intricate tasks, combining binary decision-making with arithmetic computations. Even though it is still a binary value, the output is a more thorough assessment than the first rule and is appropriate for situations when binary judgment based on intermediate values and quantitative analysis are needed.

  Using this rule, we investigate algorithm-level improvements in the context of FHE and their effects on runtime efficiency in more detail. More precisely, we are concerned with maximizing the computation of absolute value—a calculation that is frequently performed in statistical analysis. However, in certain instances, manual implementations and opti-mizations are required because certain FHE tools do not directly support certain processes.

  We examine the differences in runtime performance between various methods for calcu-lating absolute value for the TFHE-rs method. Among these methods are:

  – **Naive manual implementation:** A simple approach that uses branching to manually calculate the absolute value without any special optimizations.

  – **Optimized implementation done Manually:** An enhanced manual implementation that makes use of optimizations to boost performance without depending on native library API.

  – **Utilizing the built-in *if then else* API for implementation:** To calculate the abso- lute value, use the integrated if then else interface of TFHE-rs.

  – **Utilizing the built-in *abs* API for implementation:** Utilize the dedicated abs inter- face for absolute value computation provided by TFHE-rs.

Figure 7 illustrates the findings, which show that the manually optimized code performs at a level that is comparable to the if then else approach. But the most effective choice is undoubtedly the native Abs API, which achieves a runtime of about 0.8 seconds for regular operations, including client and server sides, without the need for a key generation pro-cedure. The abs API of TFHE-rs employs a sophisticated technique using bit operations and masks to compute absolute values without branching. [21] When compared to con-ventional branching methods, this approach is known to perform better and offer a more effective solution that is appropriate for the FHE situation.

Since the library does not have a specialized interface for directly computing absolute val-ues, we investigated a number of approaches for the OpenFHE solution and are presenting

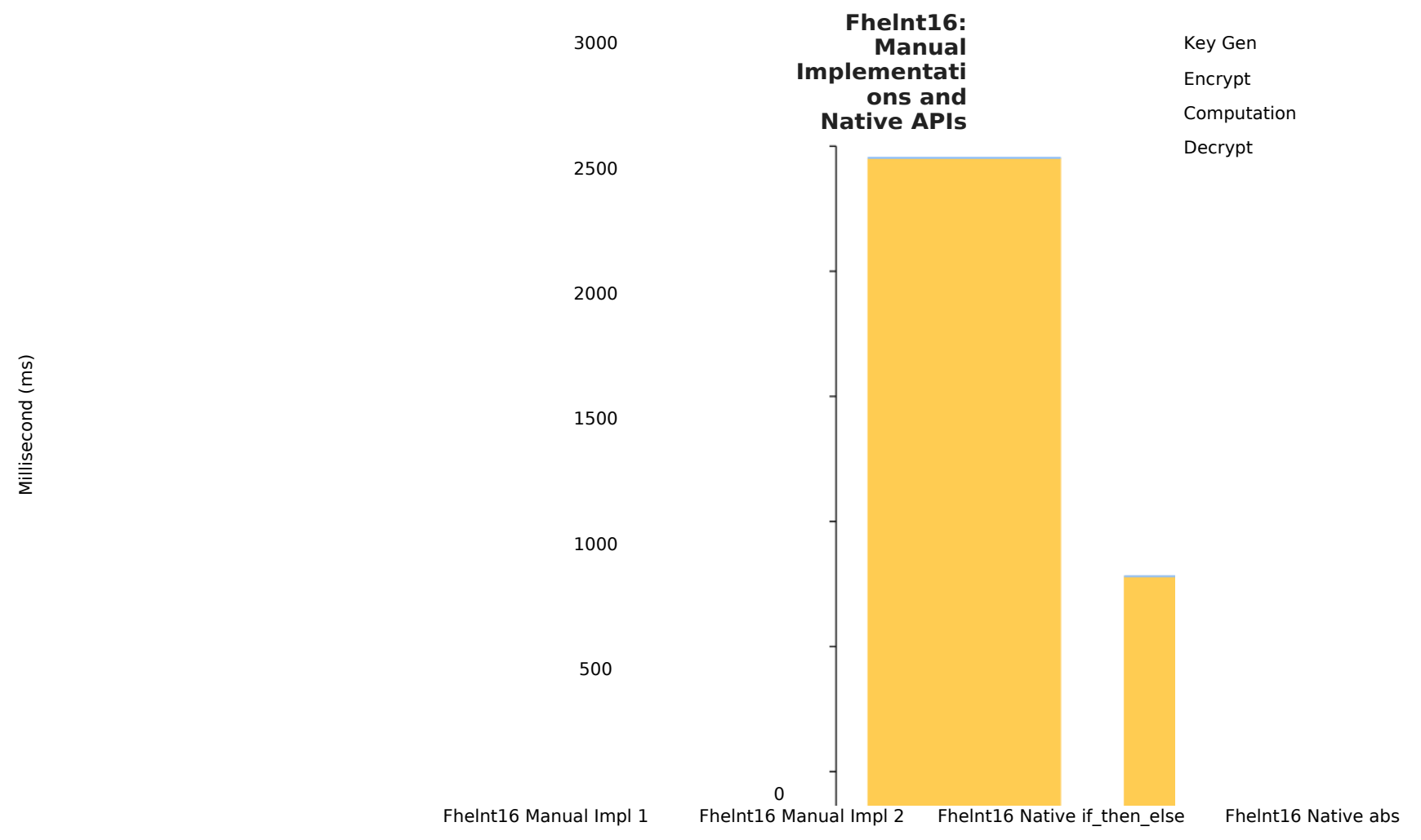

Figure 7. Top implementations of TFHE for the second QC rule, comparing the efficiency of various methods.

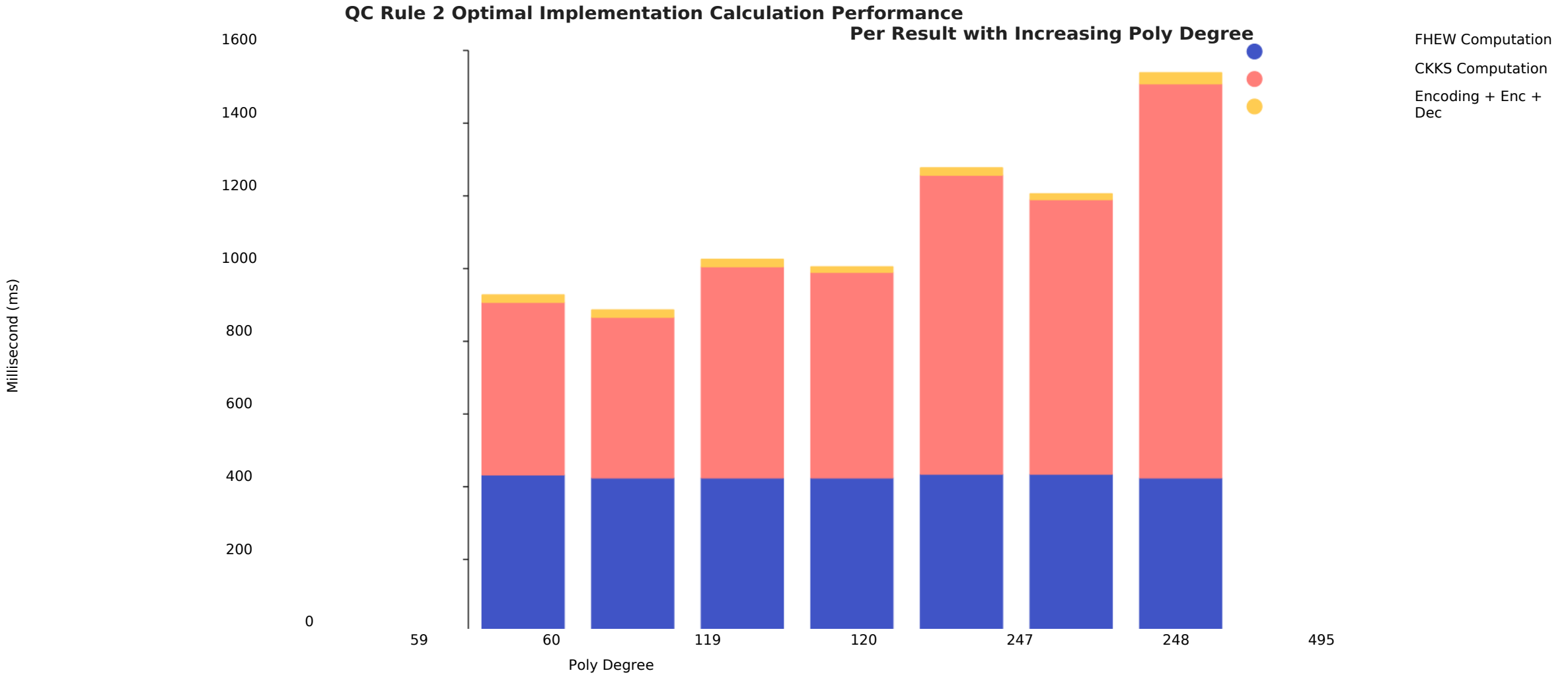

Figure 8. Performance of OpenFHE implementations with respect to the second QC rule as impacted by
polynomial degree.

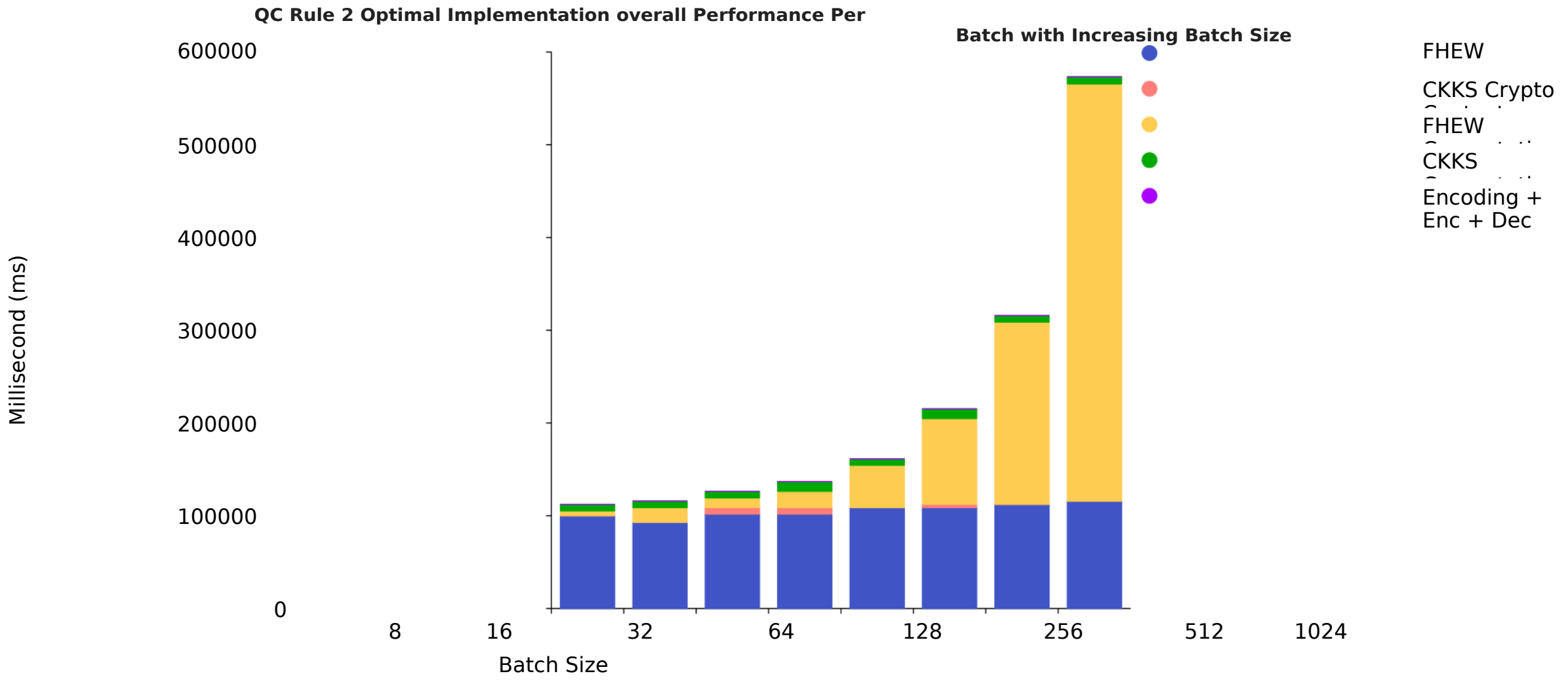

Figure 9. Implementation of the second QC rule in OpenFHE, measuring performance as batch sizes increase.

the optimal one based on runtime

putation simple. The suboptimal approach, meantime, makes use of OpenFHE's unique ability to compute any functions using Chebyshev function evaluation. This adaptability is not without its drawbacks, though, especially when choosing polynomial degrees. The degree of polynomial has a considerable impact on both accuracy and performance, as the evaluation of Chebyshev functions is essentially a polynomial approximation: a higher de-gree results in improved accuracy but at the expense of slower calculation, and vice versa. Performance and precision are brought into a crucial balance as a result. Taking this into account, the study evaluated how polynomial degree affected runtime on the client and server sides, as seen in figure 8. Next, in order to achieve the optimum performance and sufficiently accurate findings (i.e., an error rate less than 1 in 100,000), we choose an ac-ceptable polynomial degree for the given context. This prompted more testing of the run-time performance as batch sizes increased, as figure 9 illustrates. In addition, we assessed the performance of the optimal and suboptimal approaches for each result using figure 10. The suboptimal technique is only somewhat slower than the ideal way, especially at big-ger batch sizes, but it is more general in a wider range of computations, highlighting its importance even though the optimal method turned out to be more efficient.

Figure 11 shows that the complexity for setting up the cryptographic environment is in-creased by OpenFHE’s unavoidable technique switching when TFHE-rs and OpenFHE’s best implementations are compared across different batch sizes. However, when look-ing at the average daily runtime per result, the batching functionality of OpenFHE stands out as a clear benefit. OpenFHE may achieve runtimes of about 400 milliseconds with a batch size of 1024, which is almost twice as fast as those of TFHE-rs solutions, as Figure 12 illustrates. The reason for this efficiency boost is because the majority of the arith-metic computations in this QC rule are made possible by the CKKS scheme and batch processing. This explains why, from a theoretical perspective, OpenFHE exhibits superior efficiency in this specific method.

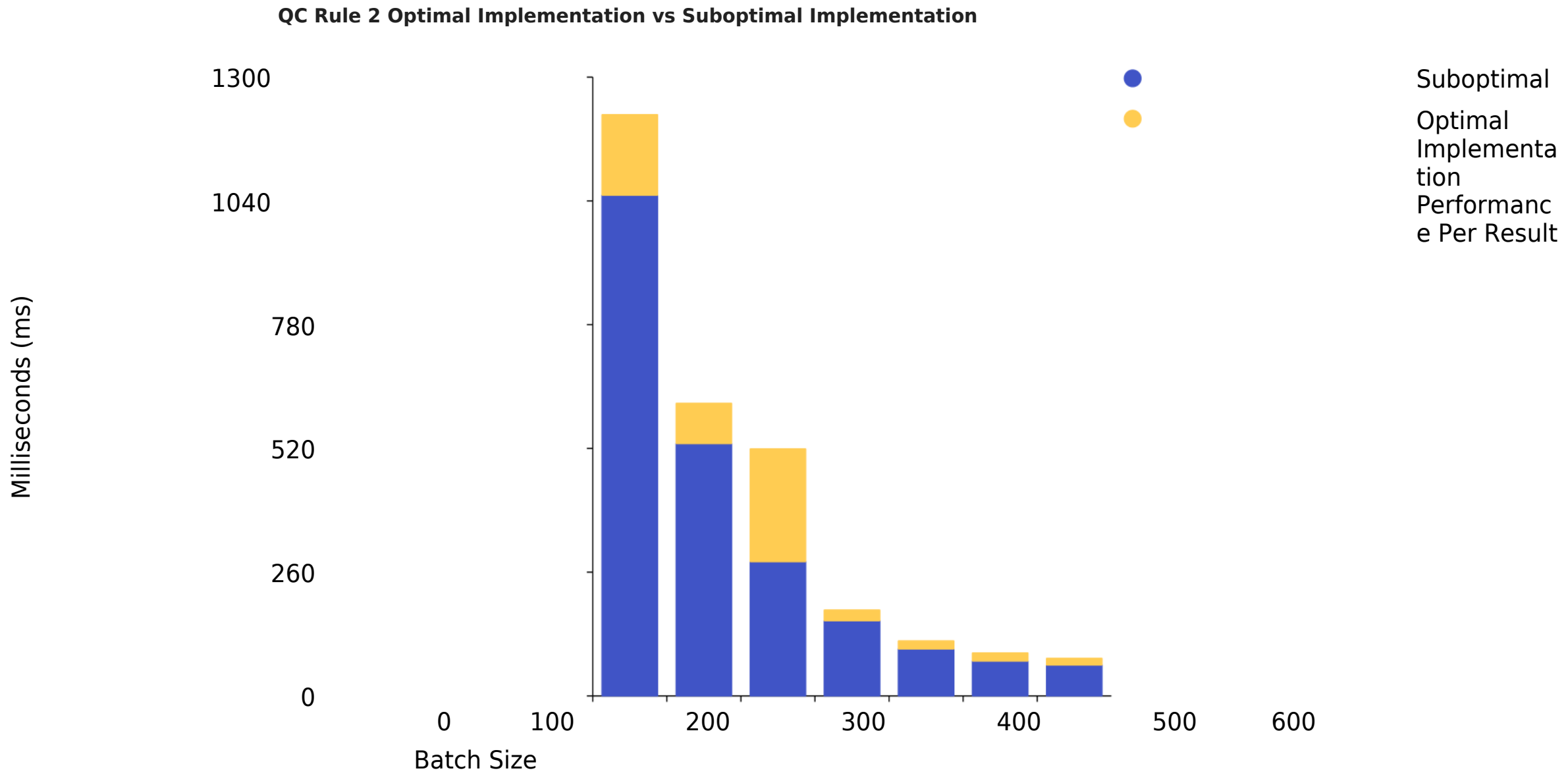

Figure 10. Implementation of the second QC rule in OpenFHE: comparing optimal and suboptimal methods for performance per result.

- **Rule 3: Sophisticated aggregation and assessment**

In this scenario, this technique is the most complicated. It uses a broad range of binary and arithmetic operations intended for in-depth examination. This algorithm functions on a matrix of decimal numbers as multidimensional data, in contrast to the preceding principles. It produces a decimal after going through several rounds of aggregation and assessment. Since this technique is made to work with different input sizes, we have accounted for this variable in the benchmarks by calling it ”input vector length.”

Here, we investigate further how parallelism and algorithm-level improvements can im-prove efficiency as input vector length increases with TFHE-rs. Rearranging certain cal-culations to make them more consistent with the FHE environment and streamlining com-putations are examples of algorithmic optimizations. Our emphasis in terms of parallelism is on computing maximum (max) and minimum (min) values simultaneously, as well as standard deviation (sd) values in parallel. Performance significantly increased as a result of these upgrades. The optimized method outperforms the naive approach by around six times, as seen in figure 13 for server-side processing. It is noteworthy that there is an in-crease in CPU utilization from 1011 percent to 1476 percent, indicating improved parallel processing skills. Taking into account that the experiment was carried out on commodity hardware with 16 threads, the most it can achieve is 1476 percent. Hardware optimized for FHE operation can function better.

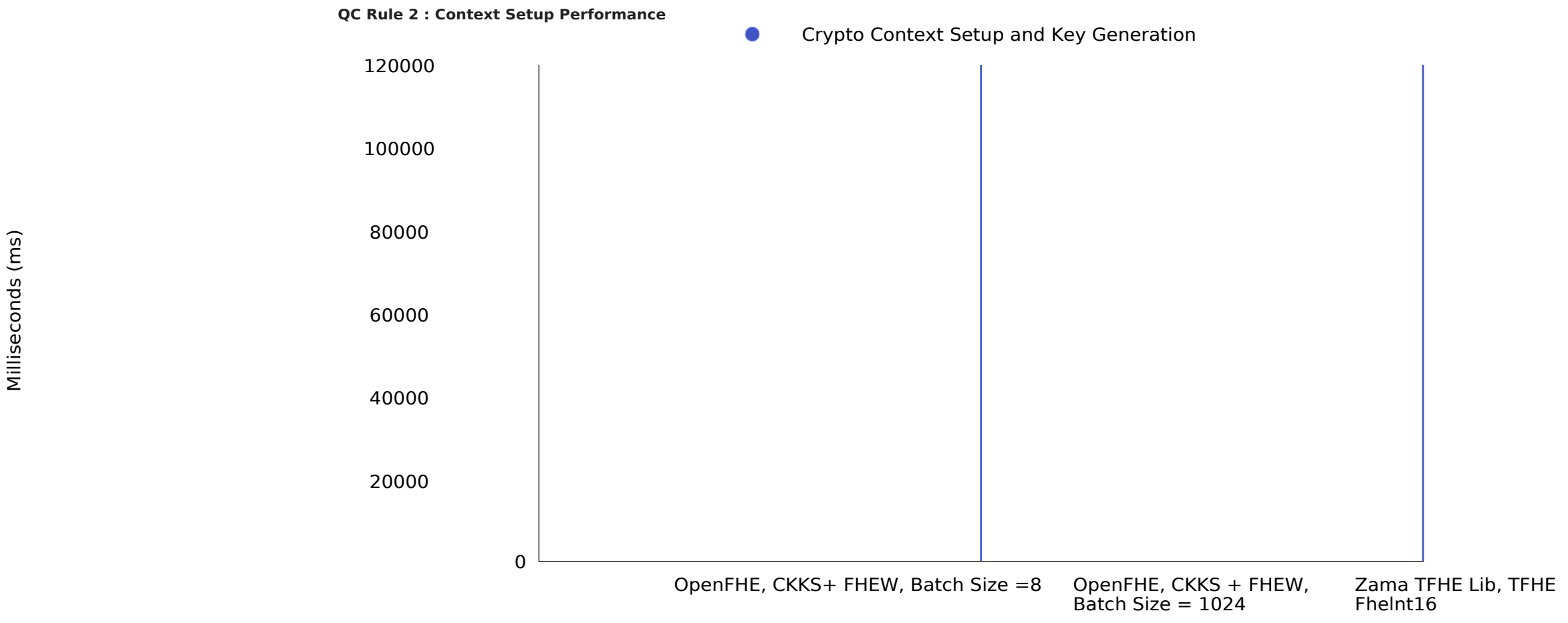

Figure 11. Contrast the second QC rule implementations in TFHE-rs and OpenFHE, raphy setup and key generation performance.

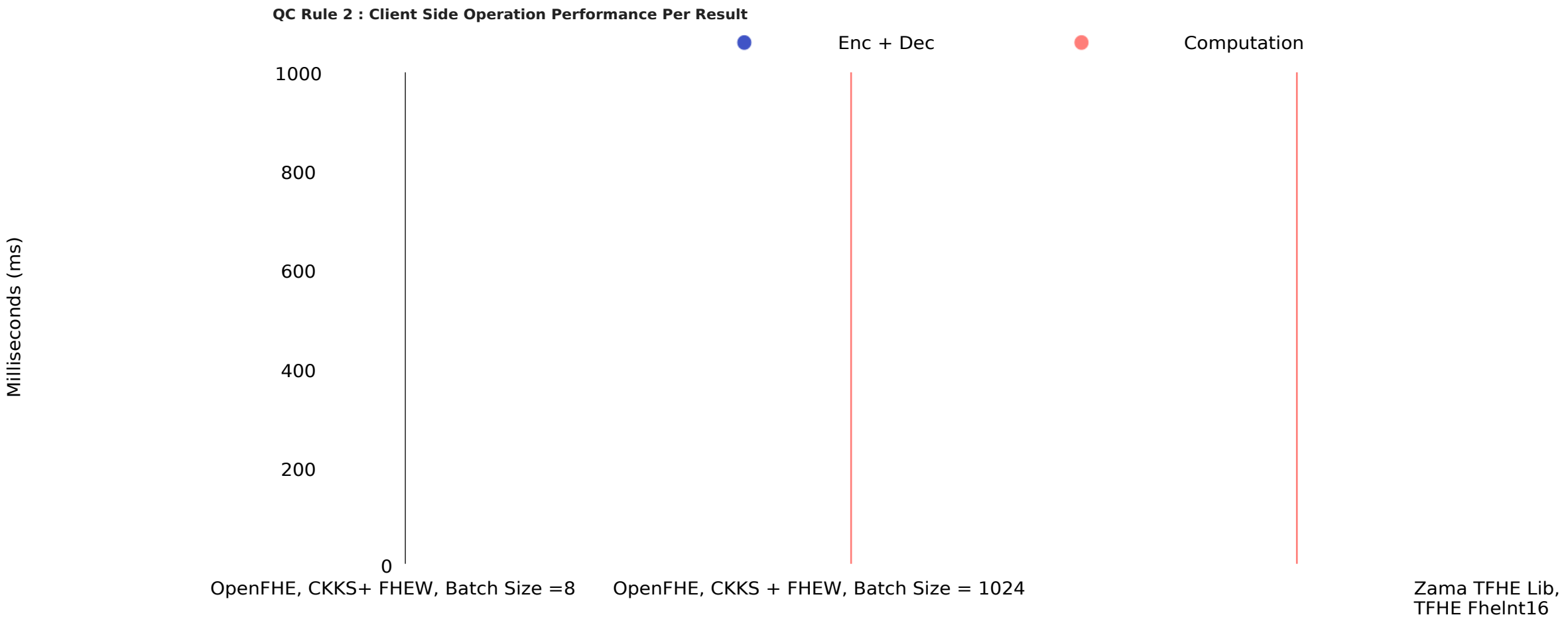

Figure 12. Contrast the second QC rule implementations in TFHE-rs and OpenFHE, operational performance on both the client and server side.

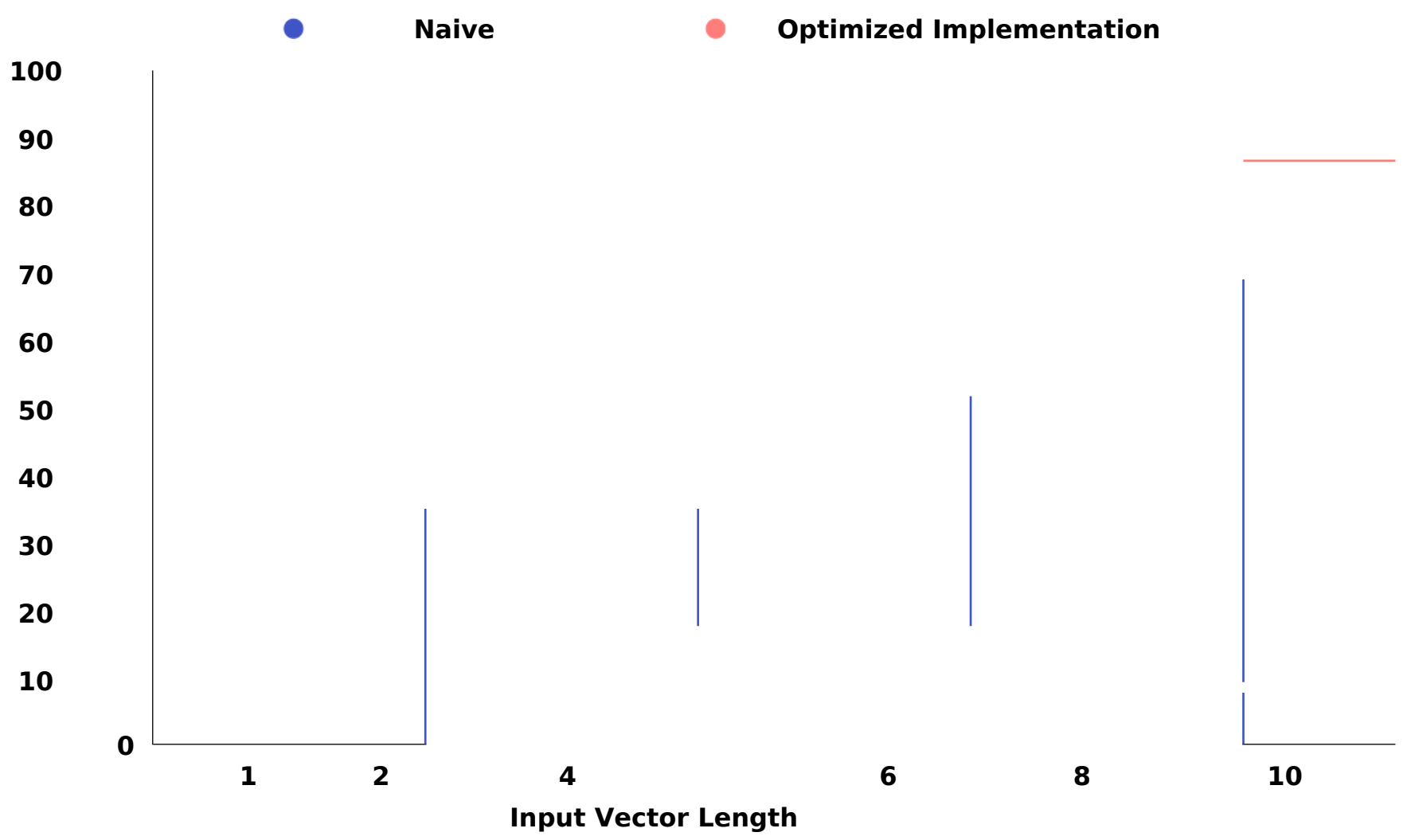

Figure 13. Implementations of the third QC rule in TFHE-rs, client-side performance improvement as batch size increases.

Figure 14 illustrates how we evaluated client performance over a range of input vector sizes. In summary, the time required for key generation is consistently less than 300 milliseconds (ms); decryption takes very little time; and encryption takes a little amount of time that increases with the number of ciphertexts.

This algorithm aggregates data for every row and every column in a matrix of inputs. The advancement of FHE using OpenFHE is significantly hampered by this grouping in two-dimensional data processing. One issue is that this tool can only do specific calculations on a complete batch of data—not on a portion or group of data. OpenFHE, for instance, is limited to executing max/min calculations within a single batch. As a result, figuring out the maximum value for each row in a matrix of decimals becomes difficult. One can obtain a maximum and minimum value for the entire matrix by encoding it in a single batch, rather than obtaining values for each row separately as would be anticipated. The real benefit of batching is eliminated when each row is encoded as a separate batch, which would still accomplish the same result but effectively reduce the process to processing multiple rows in parallel rather than multiple matrices. In this case, the batch size is just the length of an input row rather than the expected number of matrices. Another drawback for real-world applications is that experiments show that processing numerous batches in parallel would result in a large rise in runtime memory utilization.

Furthermore, OpenFHE presents an additional difficulty in terms of output correctness. Deviations result from the approximate nature of the CKKS method, and shifting to binary

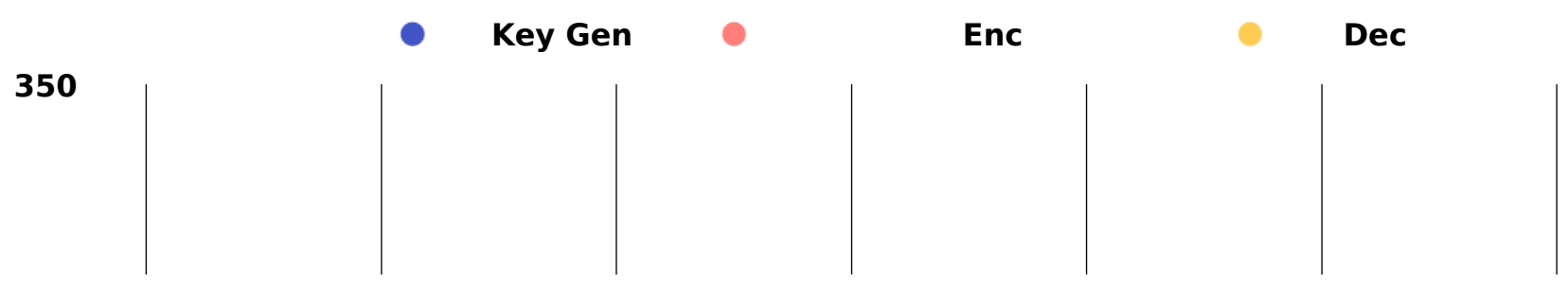
Key Gen
Enc
Dec
350

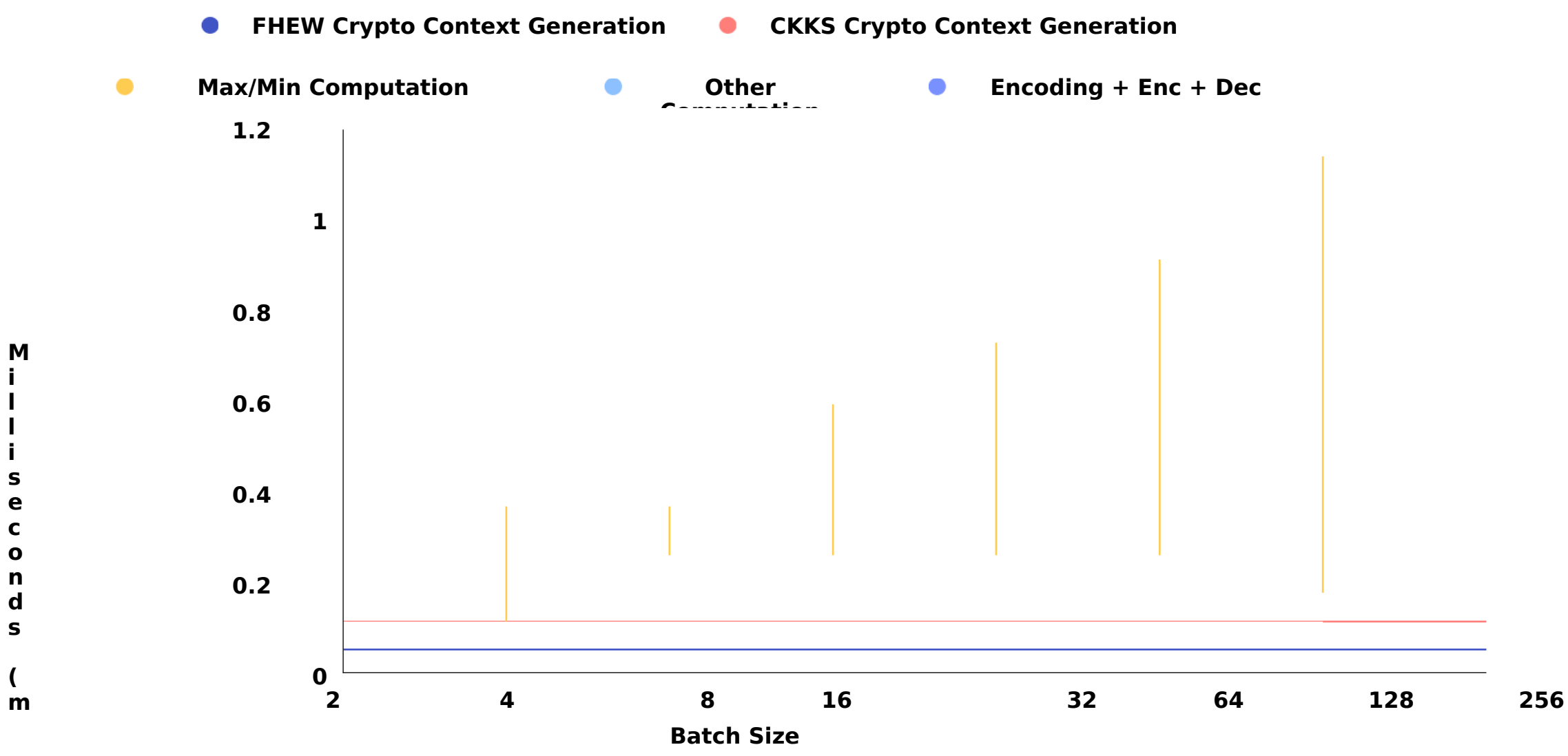

Figure 15. Implementation of the third QC rule in OpenFHE, performance evaluation as batch size increases.

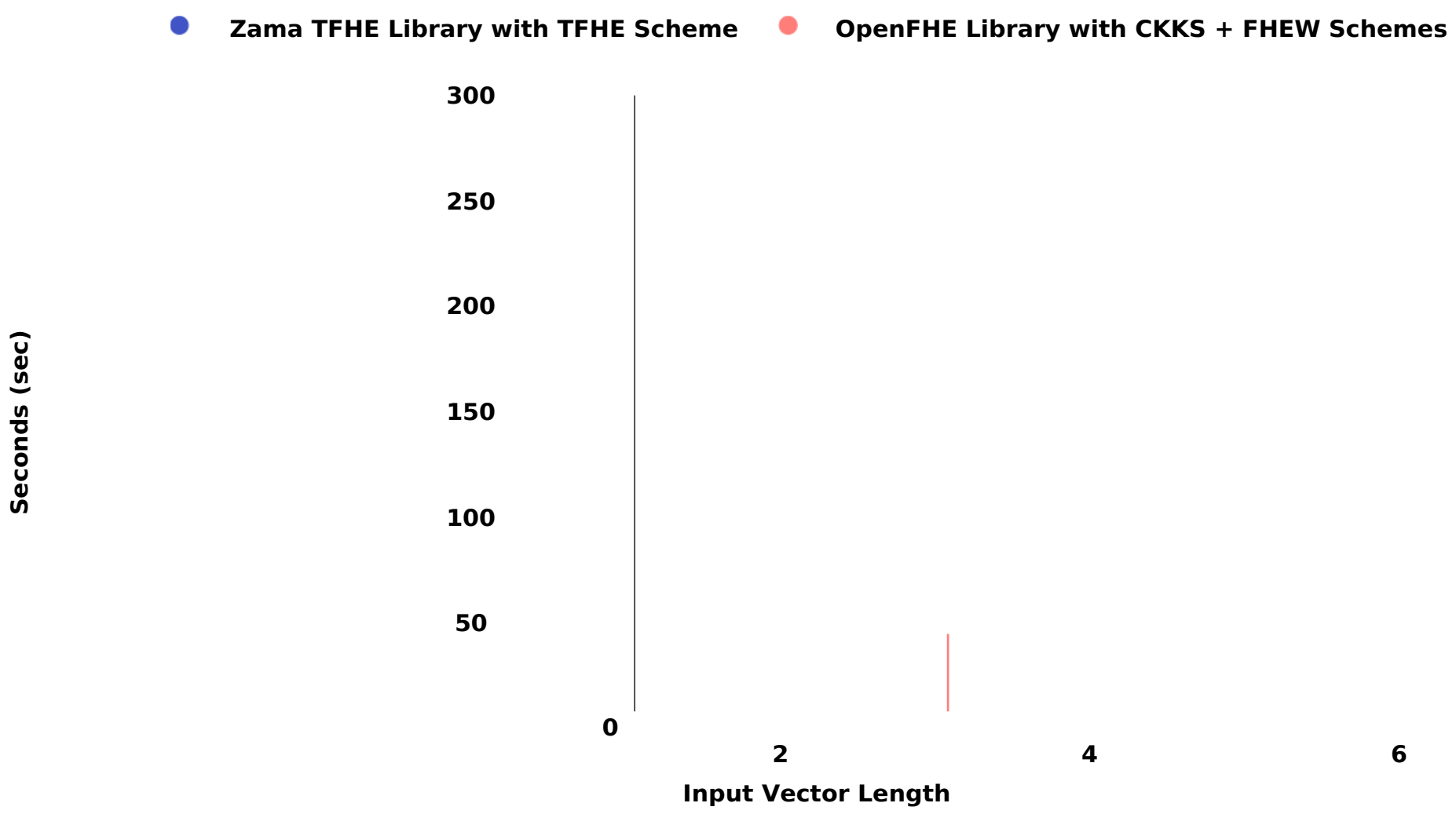

Figure 16. Contrast the third QC rule and OpenFHE.

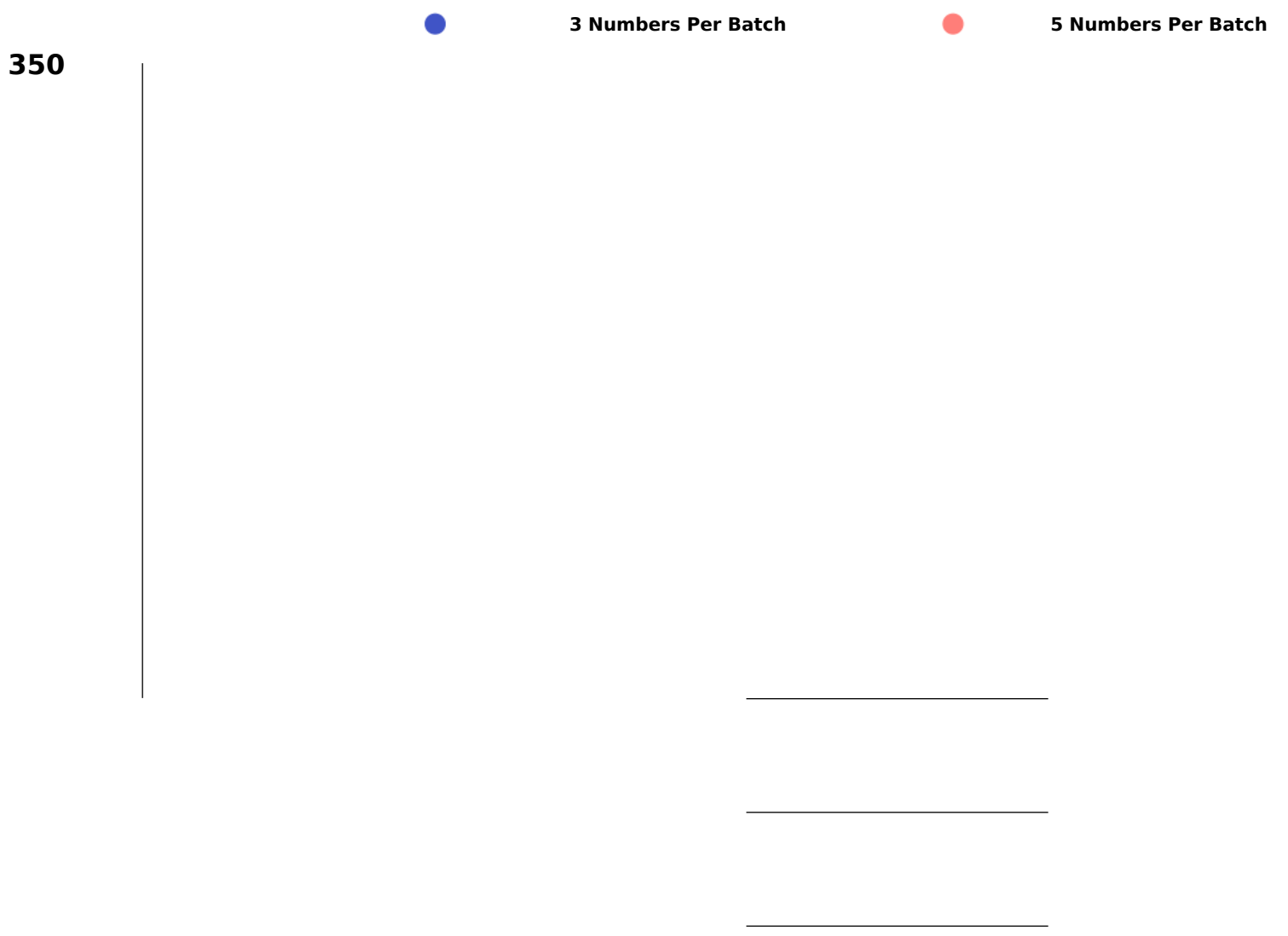
3 Numbers Per Batch
5 Numbers Per Batch
350

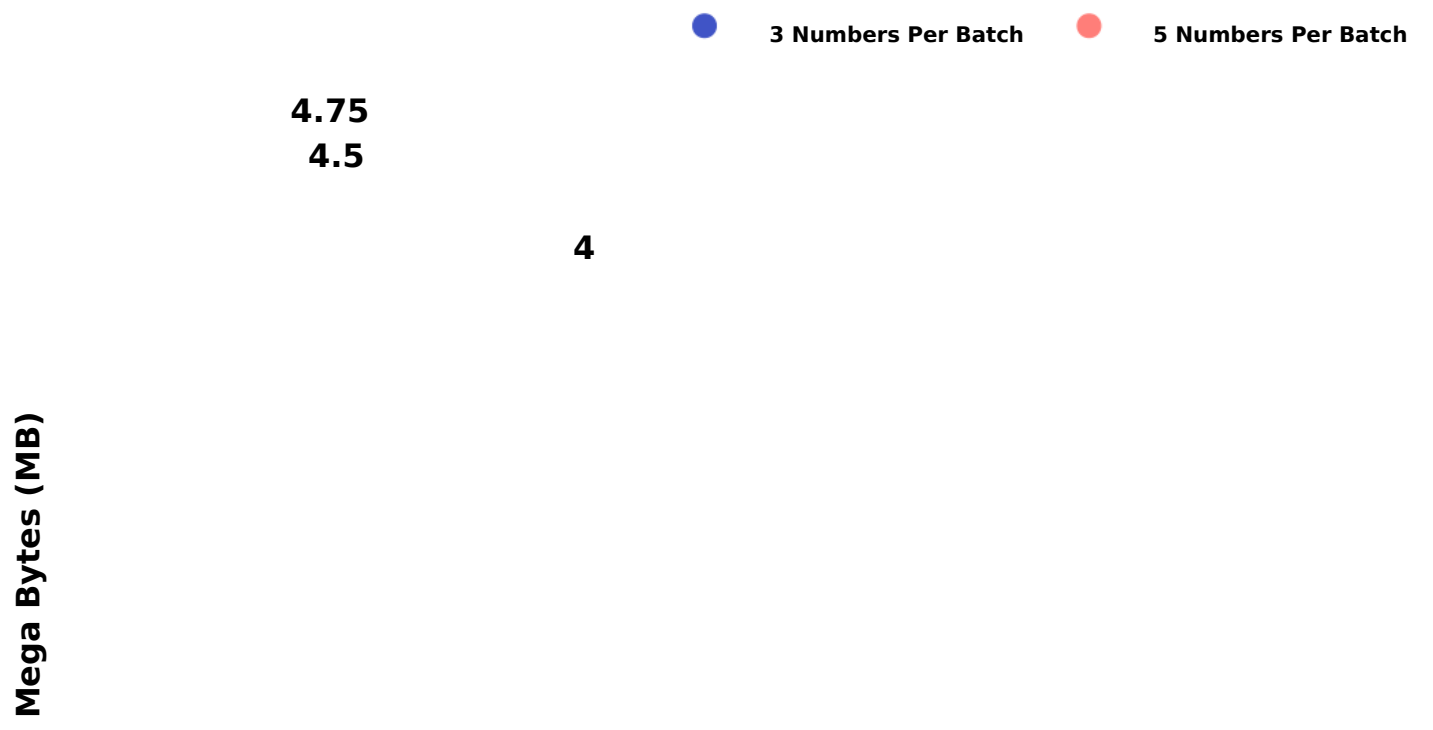
3 Numbers Per Batch
5 Numbers Per Batch
4.75
4.5
4
Mega Bytes (MB)

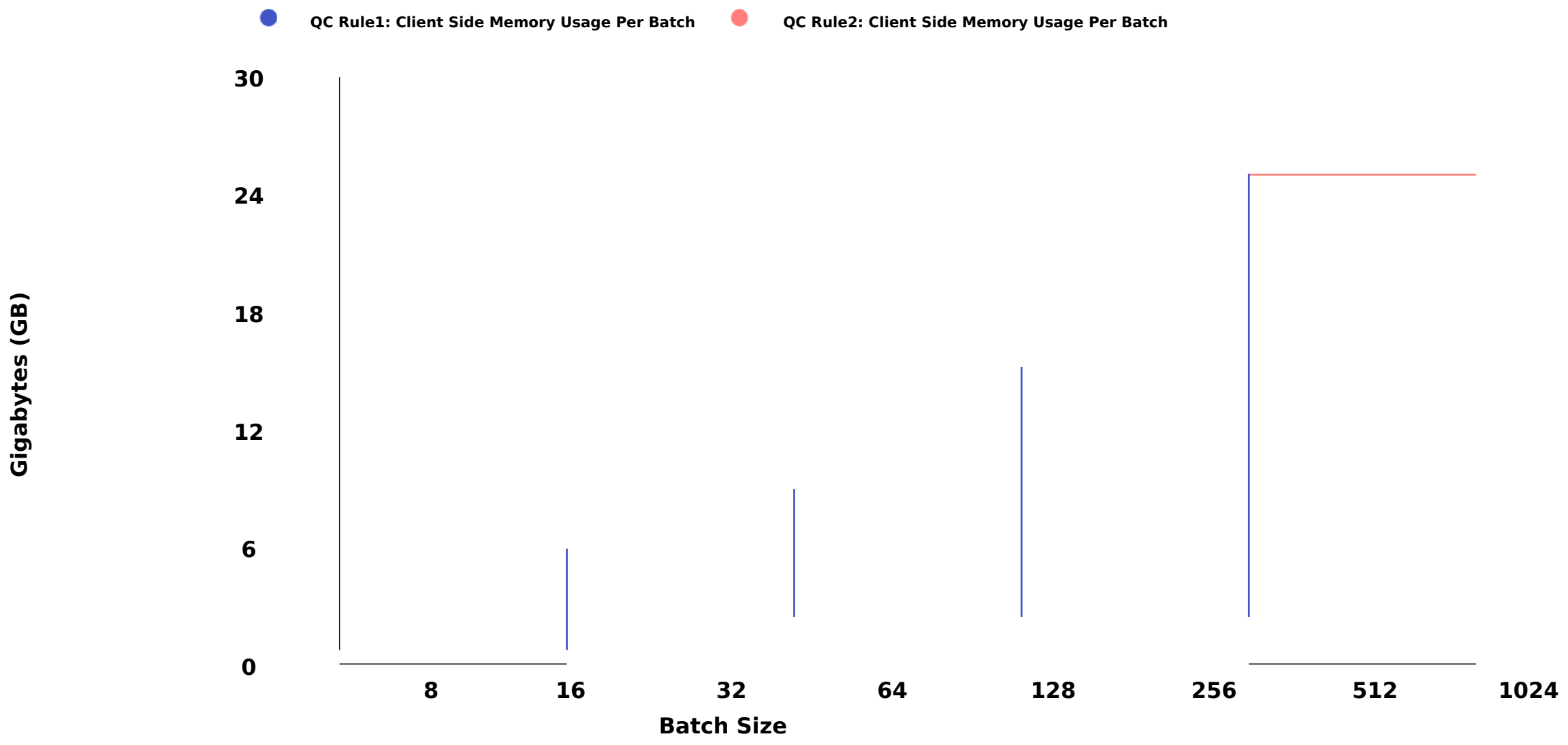

Figure 20. Memory usage at runtime on the client side for the OpenFHE implementations of algorithms.

| Key and Encrypted data | Size |
|---|---|
| CKKS CryptoContext | 1.1 KB |
| FHEW CryptoContext | 0.2 KB |
| CKKS Public Key | 18.9 MB |
| CKKS Multiplication Key | 56.6 MB |
| One Batch of CKKS Ciphertext | 13.6 MB |

Table 4. Size of keys and encrypted data for OpenFHE.

The fundamental idea behind neural network (NN) models that have been modified for Fully Homomorphic Encryption (FHE) is to allow the models to function on encrypted data in the prediction stage, while requiring original, unencrypted input for training. In delicate industries like healthcare and banking, this strategy is especially important as it guarantees data protection and privacy.

- **Methods of development :** There are presently two primary approaches to creating NN models that are interoperable with FHE: converting current NN models into equivalents for FHE and creating NN models in an environment that is native to FHE.

  The first approach can be carried out using high-level tools, which offer more abstracted, user-friendly interfaces for conversion, or low-level tools, which give more detailed control over the conversion process but demand a significant amount of engineering work. Models that have been shown to be well-designed in their unencrypted form can benefit from this strategy; nevertheless, the conversion procedure can be laborious and could result in accuracy loss due to FHE's constraints.

  As an alternative to translating from non-FHE models, the second approach, which builds NN models directly in an FHE native environment, seeks to increase inference accuracy and simplify the model-building process. An extensive range of functions resembling those in conventional ML/NN systems must be supported by FHE tools, which must be quite sophisticated under this direct approach. The capabilities of FHE tooling are severely limited by this strategy, despite the fact that it provides a cleaner integration with FHE.
- **Workflow of client-server system :** There are two stages to the process of using NN applications that are compatible with FHE. The NN model is first created by the client and sent to the server, which is situated in the trusted zones of the healthcare industry, during the preparation phase. This is appropriate in our scenario because the server uses clear, unencrypted data to train the model. After that, the client creates key pairs (which are periodically refreshed) and encrypts the inputs before delivering them to the server during the daily NN inference phase. After that, the server uses the FHE-compatible method to process the encrypted data and sends the encrypted output back to the client. To achieve the final output, the client decrypts and decodes this information. The original use case of the QC algorithm and this procedure are nearly the same.

In the context of healthcare, this research focuses on three different NN models, each with a different amount of capability and complexity. A simple categorization model is the first NN model. This model is important for understanding how current FHE tools handle simple NN models, even if it is trivial and not used in production. The second model is intended for categorization and picture analysis. The architecture of the model, a fully connected neural network (NN) with numerous layers and a huge number of parameters, is remarkably similar to production-level models that use medical image analysis to detect abnormalities and malignancies. Lastly, and by far the most complicated of the three, is the PCR test model. For the analysis of PCR tests, this approach is actively employed in production. It presents difficulties for the existing HE tooling and development because it involves a complex design with more than 40 layers and many parameters. This use case attempts to investigate the viability of using FHE techniques in practical deep learning applications for healthcare.

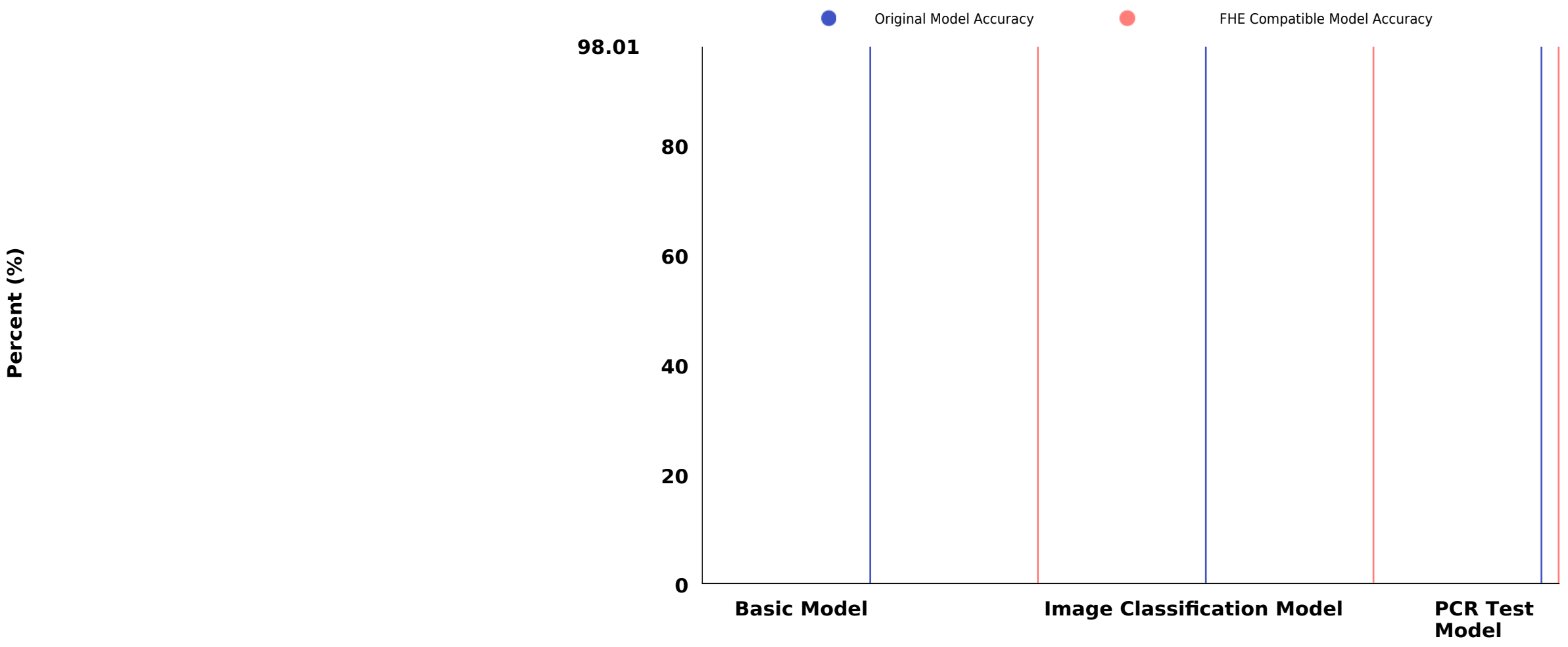

Figure 21. Accuracy of neural network inference in normal data and fully settings for the basic model, image classification model, and PCR test model,

In the research, we used the first development approach on both the PCR test model and the first basic model, which included transforming existing NN into FHE equivalents. In the meantime, we decided to use Concrete-ML's native API to create and train the second image classification model, following the second development strategy. It should be noted that the PCR test model's high degree of customization and complexity made translating it difficult. Not all operations could be directly translated into FHE through low-level interfaces, and not all operations could be supported by high-level interfaces. This resulted in several major changes to the model, such as the removal of several preprocessing layers, a modification to the input shape, and the replacement of some unsupported operations with tooling-supported ones.

The inference accuracy of these models is compared between clear data and FHE settings in Figure 21. Because it was so basic, the first basic model was still accurate after FHE conversion. Remarkably, there was a minor drop in inference accuracy for the second model, which was created in an FHE native context. Conversely, the third PCR test model had a discernible drop in accuracy, which may have been brought on by post-training quantization, a procedure known to have an impact on accuracy. Furthermore, it is possible that the model's adjustments played a role in the recorded decline in accuracy.

The inference performance of the three models in the framework of FHE is shown in Fig-ure 22. While the first basic model may finish an inference in less than a second, the PCR test model and the more complicated image classification model take much longer—about 20 or 30 seconds—to accomplish a single inference. This means that one inference takes

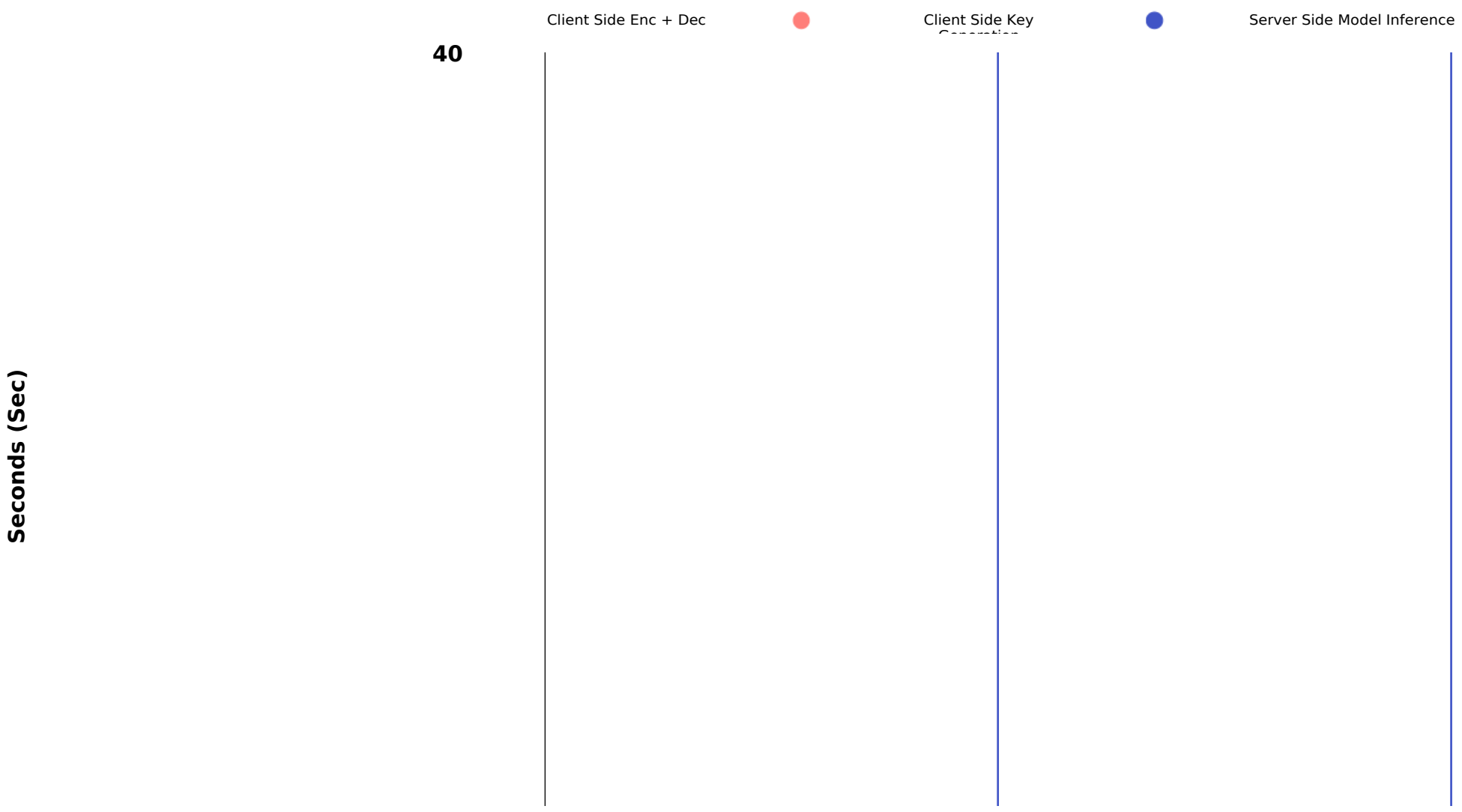
Client Side Enc + Dec
Client Side Key
Server Side Model Inference
40
Seconds (Sec)

into the FHE setting, albeit at the expense of extra transmission expenses, calculation overhead, and storage needs. Even with these developments, NN models meant for production still need to overcome a number of obstacles before they can be considered ready for FHE integration. These models have significant bottlenecks due to their intrinsic complexity.

**Performance and resource utilization :** The best results obtained in the QC use case are deemed sufficient for applications when prompt responses are not essential. While certain non-latency-sensitive jobs can be completed with the achieved performance, more advancements are obviously required to increase effectiveness and broaden application.

This isn't the case with NN models. For most applications, a performance of roughly 30 seconds per inference is far from realistic, particularly in situations where millions of inferences are expected to be generated daily. Across all NN models, inference accuracy continues to be a major challenge. A small accuracy drop of about 0.5% might be acceptable for some applications, particularly if it means receiving the advantages of FHE. Nonetheless, there are applications where great accuracy is required and cannot be compromised, especially when using complicated NN models for essential assessments. Furthermore, the input size for NN applications connected to healthcare might be very huge, such 4K pictures. The size of these inputs can grow even more significantly after encryption, posing serious storage and communication problems. These factors present significant obstacles to the state of the art in FHE research.

**Creating Neural Networks Compliant with FHE :** The degree of development complexity in an FHE framework is significantly influenced by the architecture of a neural network model. It is easy to construct or adapt basic models to the FHE scenario. However, to avoid the laborious translation procedure and take full use of FHE, it is advised to begin the development process with FHE native libraries and tools for highly customized or intrinsically complex models.

**Connecting research and industry to bridge the gap:** We've found a vast unmet need for FHE in statistical analysis, even though most of the necessary procedures can be accomplished effectively with current FHE tools in the context of QC use cases. Industry development might be considerably aided by filling the gap that exists between the lack of direct support for common statistical primitives like standard deviation in FHE tools now available. Enhancing performance, lowering engineering efforts, and reducing security risks are just a few of the major benefits of providing library support for these statistical procedures.

Similar to this, there is a significant disconnect between research potential and commercial appli-cation needs, even while existing FHE tools are good at handling basic and built-in NN models. Trans-lating customized models—which are common in production environments—into an FHE framework can be particularly difficult because of their complicated architecture and unique configurations. This procedure demonstrates the enormous potential for advancement in FHE tools and automation, as does the want for increased performance and the difficulty of handling big key and ciphertext sizes.

Furthermore, the industry is showing a noteworthy interest in FHE-specific hardware. The uti-lization of such hardware has the potential to greatly accelerate FHE operations, making FHE more practical and effective for a wide range of industry applications, from edge devices to cloud comput-ing.

In summary, while the incorporation of FHE into QC algorithmic processing exhibits encourag-ing outcomes, the use of FHE in NN models—particularly those intended for production—remains a developed field that requires additional work. Innovation in technology alone won't be enough to overcome current obstacles; a paradigm shift in the way models are designed and built for cryptography contexts is also necessary.

# References

[1] Rauthan J. Homomorphic encryption approach for exploration of sensitive information retrieval. *Journal of Intelligent & Fuzzy Systems*, 2020. **38**(5):6495–6505.

[2] Rauthan JS, Vaisla KS. Privacy and Security of User's Sensitive Data: A Viable Analysis. In: Proceedings of the Second International Conference on Research in Intelligent and Computing in Engineering, volume 10 of *Annals of Computer Science and Information Systems*. PTI, 2017 pp. 67–71. doi:10.15439/2017R45. URL http://dx.doi.org/10.15439/2017R45.

[3] Ji Z, Lipton ZC, Elkan C. Differential Privacy and Machine Learning: a Survey and Review, 2014. 1412.7584, URL https://arxiv.org/abs/1412.7584.

[4] Rauthan JS, Vaisla KS. Scrambled database with encrypted query processing: CryptDB a computational analysis. In: 2017 1st International Conference on Intelligent Systems and Information Management (ICISIM). 2017 pp. 199–211. doi:10.1109/ICISIM.2017.8122174.

[5] Lindell Y. Secure Multiparty Computation (MPC). Cryptology ePrint Archive, Paper 2020/300, 2020. doi:10.1145/3387108. https://eprint.iacr.org/2020/300, URL https://eprint.iacr.org/2020/300.

[6] Ernstberger J, Chaliasos S, Zhou L, Jovanovic P, Gervais A. Do You Need a Zero Knowledge Proof? Cryptology ePrint Archive, Paper 2024/050, 2024. https://eprint.iacr.org/2024/050, URL https://eprint.iacr.org/2024/050.

[7] Sabt M, Achemlal M, Bouabdallah A. Trusted Execution Environment: What It is, and What It is Not. In: 2015 IEEE Trustcom/BigDataSE/ISPA, volume 1. 2015 pp. 57–64. doi:10.1109/Trustcom.2015.357.

[8] Rauthan J. Fully homomorphic encryption: A case study. *Journal of Intelligent & Fuzzy Systems*, 2022. **43**(6):8417–8437.

[9] Iezzi M. Practical Privacy-Preserving Data Science With Homomorphic Encryption: An Overview. In: 2020 IEEE International Conference on Big Data (Big Data). 2020 pp. 3979–3988. doi:10.1109/BigData50022.2020.9377989.

[10] Munjal K, Bhatia R. A systematic review of homomorphic encryption and its contributions in healthcare industry. *Complex Intelligent Systems*, 2022. **9**:1–28. doi:10.1007/s40747-022-00756-z.

[11] Rauthan J, Vaisla K. VRS-DB: preserve confidentiality of users' data using encryption approach. *Digital Communications and Networks*, 2021. **7**(1):62–71. doi:https://doi.org/10.1016/j.dcan.2019.08.001. URL https://www.sciencedirect.com/science/article/pii/S235286481930001X.

[12] Rauthan JS, Singh Vaisla K. VRS-DB: Computation Exploration on Encrypted Database. In: 2019 International Conference on Big Data and Computational Intelligence (ICBDCI). 2019 pp. 1–6. doi: 10.1109/ICBDCI.2019.8686098.

[13] Yousuf H, Lahzi M, Salloum SA, Shaalan K. Systematic Review on Fully Homomorphic Encryption Scheme and Its Application, pp. 537–551. Springer International Publishing, Cham. ISBN 978-3-030-47411-9, 2021. doi:10.1007/978-3-030-47411-929. URL https://doi.org/10.1007/978-3-030-47411-9_29.

[14] Pulido-Gaytan LB, Tchernykh A, Cort´es-Mendoza JM, Babenko M, Radchenko G. A Survey on Privacy-Preserving Machine Learning with Fully Homomorphic Encryption. In: Nesmachnow S, Castro H, Tch-ernykh A (eds.), High Performance Computing. Springer International Publishing, Cham, 2021 pp. 115–129.

[15] Armknecht F, Boyd C, Carr C, Gjøsteen K, J¨aschke A, Reuter CA, Strand M. A Guide to Fully Homo-morphic Encryption. Cryptology ePrint Archive, Paper 2015/1192, 2015. https://eprint.iacr.org/ 2015/1192, URL https://eprint.iacr.org/2015/1192.

[16] Viand A, Jattke P, Hithnawi A. SoK: Fully Homomorphic Encryption Compilers. In: 2021 IEEE Symposium on Security and Privacy (SP). IEEE, 2021 doi:10.1109/sp40001.2021.00068. URL http: //dx.doi.org/10.1109/SP40001.2021.00068.

[17] Badawi AA, Alexandru A, Bates J, Bergamaschi F, Cousins DB, Erabelli S, Genise N, Halevi S, Hunt H, Kim A, Lee Y, Liu Z, Micciancio D, Pascoe C, Polyakov Y, Quah I, RV S, Rohloff K, Saylor J, Suponitsky D, Triplett M, Vaikuntanathan V, Zucca V. OpenFHE: Open-Source Fully Homomorphic Encryption Library. Cryptology ePrint Archive, Paper 2022/915, 2022. https://eprint.iacr.org/2022/915, URL https://eprint.iacr.org/2022/915.

[18] Zama. TFHE-rs: A Pure Rust Implementation of the TFHE Scheme for Boolean and Integer Arithmetics Over Encrypted Data, 2022. https://docs.zama.ai/tfhe-rs.

[19] Zama. Concrete ML: a Privacy-Preserving Machine Learning Library using Fully Homomorphic Encryp-tion for Data Scientists, 2022. https://docs.zama.ai/concrete-ml.

[20] James Westgard, ”Westgard Rules” and Multirule Quality Control, 2021. URL https://westgard. com/westgard-rules.

[21] Sean Eron Anderson, Bit twiddling hacks, compute the integer absolute value (abs) without branching, 2005. URL https://graphics.stanford.edu/~seander/bithacks.html.